\documentclass[twocolumn,showpacs,preprintnumbers,amsmath,amssymb,aps,prb,floatfix,groupedaddress]{revtex4}
\usepackage{graphicx}
\usepackage{dcolumn}
\usepackage{bm}
\usepackage{bigints} 
\usepackage{bm} 
\usepackage{epsfig}
\usepackage{color}
\usepackage{multirow}
\usepackage{amsmath}
\usepackage{physics}
\usepackage{tikz-feynman}
\begin{document}
\title{Two Dimensional Polarons with Linearly Dispersing Self Energy and Other Novel Features in the Study of Bi/Single Layer Graphene, and Semi-Dirac Semi-metals on Polar Substrates}

\author{S. Banerjee}
\affiliation{C2 Education, Millbrae, California, USA}

\date{\today}

\begin{abstract}
We consider the polaron dynamics driven by Fro{\"h}lich type, long wavelength dominated electron-phonon interaction, for three different semi-metals: single and bilayer graphene, and semi-Dirac, all grown on polar substrates such as, $SiC$  or $SiO_{2}$. The problem of polaron has been studied by Feynman and others for ordinary polar crystals. But the study of polaron formation in the context of the above-mentioned 2D semi-metals having non-scalar effective Hamiltonians is novel. Single layer graphene (henceforth called SL graphene), bilayer graphene (henceforth called BL graphene), and semi-Dirac have two dimensional band-structures with point Fermi surfaces in their natural undoped conditions. When SL and BL graphene are grown on polar substrates, their electrons can interact with the surface phonons of those polar substrates, as has been discussed by Fratini et al. That gives rise to the possibility of polaron formation in the context of SL and BL graphene, although they themselves are non-polar. semi-Dirac materials, which drew research interest in recent years due to is anisotropic, exotic electronic band-structure dispersing linearly in one direction and quadratically in the orthogonal direction in the Brillouin zone, was first discovered computationally in oxide heterostructures by Pardo and Pickett. In the current paper semi-Dirac, like SL and BL graphene, has been considered to be be grown on a polar substrate and the resulting polaron-dynamics has been investigated. As was discovered by Pardo and Pickett, the interfaces of $(TiO_2)_5/(VO_2)_3$ heterostructure, in which semi-Dirac dispersion was observed, are non-polar. This justifies the treatment of semi-Dirac, for the purpose of this paper, in the same footing as non-polar materials like SL and BL graphene. Staring from the Fro{\"h}lich type electron-phonon interaction Hamiltonian, second-order perturbation theory is employed to obtain expressions for the self energy of the electron due to polaron formation for the three aforementioned systems. The electron self energy, or polaron energy, calculated analytically for BL graphene for small electron-momentum, is shown to vary linearly with the electron momentum. Despite the similarity between BL graphene and ordinary polar crystals in the parabolic nature of the electronic band-structure in the absence of electron-phonon interaction, the linear energy-momentum dispersion of BL graphene polarons stands in stark contrast to the quadratic energy-momentum dispersion of the polarons produced in ordinary polar crystals. The appropriate ranges of values of the electron momenta for the formation of polarons, are established for SL and BL graphene and semi-Dirac. Once the electron momentum exceeds such an upper-bound, the energy of electron dissipates by creating a phonon. The rate of this dissipation or decay process (the decay rate) is calculated for the three systems.          
\end{abstract}

\maketitle

\section{Introduction}
There has been extensive studies on the electron-phonon interaction in the context of SL and BL graphene. In those studies, analytical expressions have been derived for the electron-phonon interaction Hamiltonians, considering both the acoustic and the optical modes of phonon-vibrations \cite{GraphehePhononAcoustic1, GraphehePhononAcoustic2, GraphenePhononOptical1, GraphenePhononOptical2, BilayerGPhononOptical}. But in all of them, the electron-phonon interaction isn't dominated by long wavelength phonons. 
\\
\\
Long wavelength phonons play an important role in the electron-phonon interaction in polar crystals. In such materials an optical phonon mode, generated by the oppositely charged ions in an unit cell moving towards each other, is accompanied by polarization (dipole moment per unit area/volume) $\vec{P}$. The electrostatic potential resulting from such polarization modifies the energy of a nearby electron. This is the origin of the Fro{\"h}lich type electron-phonon interaction. As it turns out to be the case, in such an interaction the electron-phonon coupling strength, a phonon-wavelength-dependent factor, becomes very strong in the long-wavelength region. In fact, the square of the coupling strength behaves like the electrostatic Coulomb potential\cite{2DPol3} in the Fourier space, i.e., $\vec{q}$-space, $\vec{q}$ being the wave-vector. Like the Coulomb potential, with $\abs{\vec{q}}$ the square of the coupling strength varies as $\frac{1}{\abs{\vec{q}}}$ and $\frac{1}{\abs{\vec{q}}^2}$, for a 2-D and a 3-D system respectively\cite{2DPol3, Feynman}, becoming very large for small $\abs{\vec{q}}$ or long wavelength. This type of long-wavelength dominated electron-phonon interaction, which is at the heart of Fro{\"h}lich interaction, is the cause of polaron formation in polar crystals. Fro{\"h}lich type of interactions play a dominant role at and near the interfaces of hetero-structures\cite{Mahan, Ando}. For example, when one considers a sheet of graphene (single or bi-layer) being placed on a polar substrate like $SiC$, the optical phonons at the surface of the latter can couple to the electrons of the former through the above-mentioned interaction. Such a scenario has been considered in a paper by Fratini et. al.\cite{PolGAndBLG}, in which the primary interest was to study thermodynamic properties like the resistivity of the material. 
\\
\\
In this paper we will be using the same Fro{\"h}lich type electron-phonon interaction Hamiltonian as appears in [\onlinecite{PolGAndBLG}], for our study of polaron formation in SL, BL graphene and semi-Dirac material grown on polar substrates. We will calculate the electron self energy, or polaron energy as it is also called, using the above-mentioned Hamiltonian for the range of the electron-momentum in which polaron formation is possible. We will also consider the decay rate when the electron momentum is beyond that range. The organization of the paper is as follows. Having introduced the electron-phonon interaction Hamiltonian, the polaron formation in  BL graphene on a polar substrate is considered first. After calculating the polaron energy as well as the decay rate for BL graphene, we next study the polaron dynamics of SL graphene, and finally of semi-Dirac, both materials being considered on polar substrates. In both cases we investigate the polaron energy and the decay rate. The reason for treating BL graphene first is that, unlike SL graphene and semi-Dirac, it has been possible to obtain an analytical result for the polaron energy of BL graphene for small values of the electron momentum. This analytical result clearly shows that there exists a linear relationship between the polaron energy and the corresponding momentum, when the momentum is small.

\section{Fro{\"h}lich type electron-phonon Interaction Hamiltonian}
We briefly explained the origin of the Fro{\"h}lich interaction in the introduction section. In the following, we will first write the Hamiltonian of such an interaction. For our problem the interaction takes place between the electrons of the material, e.g., BL graphene sheet, and the surface phonons of the substrate the sheet is placed upon. We will then explain various components of the Hamiltonian. The Hamiltonian is as follows\cite{PolGAndBLG}.    

\begin{equation}
\begin{aligned}
H = \Sigma_{\vec{q}}M_{\vec{q}} \rho_{\vec{q}} (b_{\vec{q}} + b^{\dagger}_{-\vec{q}})
\label{eq:PolaronHamil}
\end{aligned}
\end{equation}
In Eq.~\ref{eq:PolaronHamil}, the second quantized Bosonic operators $b_{\vec{q}}$ and $b^{\dagger}_{-\vec{q}}$ correspond to the annihilation and creation of an optical phonon. They satisfy the standard Bosonic commutation relations, viz.,$[b_{\vec{k}}, {b^{\dagger}_{\vec{q}}}] = \delta_{\vec{k}, \vec{q}}$, and $[b_{\vec{k}}, {b_{\vec{q}}}] = 0$. The frequency of an optical phonon is independent of its wave-vector. 
\\
\\
$\rho_{\vec{q}}$ in Eq.~\ref{eq:PolaronHamil} is the Fourier transform of the second quantized density operator $\Psi^{\dagger}(x)\Psi(x)$ for the electrons\cite{Kittel}, where $\Psi(x)$ is the second quantized version of the real space eigen-function for the `non-interacting' or `free' Hamiltonian of the material. [`non-interacting' or `free' in the sense that the Hamiltonian has no electron-phonon interaction]. For SL and BL graphene and semi-Dirac, the `non-interacting' Hamiltonians are matrices. As will be shown next, this results in the second quantized electron density operator appearing in Eq.~\ref{eq:PolaronHamil} having eigen-spinors in it. The eigen-spinors correspond to the pseudospin degree of freedom and have nothing to do with the actual spin. The electronic wave-function corresponding to the `non-interacting' Hamiltonian is written as $\Psi(x) = \frac{1}{\sqrt{A}}\sum_{\vec{p}}u_{\vec{p}}e^{i\vec{p}\cdot \vec{x}}c_{\vec{p}}$, $A$ being the physical area of the system. $A$ is used for normalizing the wave-function $\Psi(x)$. $u_{\vec{p}}$ is the eigen-spinor corresponding to the `non-interacting' Hamiltonian in the momentum space, and $c_{\vec{p}}$ is the second quantized Fermionic operator, which corresponds to the annihilation operator of an electron of wave-vector $\vec{p}$. We will ignore the mention of actual spin (not pseudospin) explicitly to keep notations simple. The effect of the real spin will be incorporated by inserting the spin degeneracy factor $g_s$ in the final expressions for polaron energy and the decay rate\cite{Feynman}. The operators  $c$ and $c^{\dagger}$ satisfy the standard Fermionic anti-commutation relations, viz., $\{c_{\vec{p}}, c_{\vec{l}}^{\dagger} \} = \delta_{\vec{p}, \vec{l}}$, and $\{c_{\vec{p}}, c_{\vec{l}} \} = 0$. Using the above-mentioned expression for $\Psi(x)$, $\rho_{\vec{q}}$, the Fourier transform of the electron density function $\Psi^{\dagger}(x)\Psi(x)$, defined as $\int dx \Psi^{\dagger}(x)\Psi(x) e^{i\vec{q} \cdot {\vec{x}}}$, assumes the following form
\begin{eqnarray}
\rho_{\vec{q}} = \Sigma_{\vec{k}} u^{\dagger}_{\vec{k}+\vec{q}}u_{\vec{k}}  c^{\dagger}_{\vec{k}+\vec{q}}c_{\vec{k}} 
\label{eq:ElecDens}
\end{eqnarray}
To derive the above expression for $\rho_{\vec{q}}$, the following identity, encountered frequently in Quantum field theory derivations,  is used. The identity is: $\int d^2\vec{x} e^{i(\vec{p}-\vec{p^\prime}).\vec{x}}=A\delta_{\vec{p}, \vec{p^\prime}}$, A being the physical area. It is noted that $\rho_{\vec{q}}$ in Eq.~\ref{eq:ElecDens} contains the eigen-spinors as was mentioned before. 
\\
\\
The long-wavelength (small $\abs{\vec{q}}$) dominated term $M_{\vec{q}}$ in Eq.~\ref{eq:PolaronHamil} is given by ${M_{\vec{q}}}^2 = g\frac{e^{-2qD}}{q}$, where $q$ is $\abs{\vec{q}}$. $g$ is the substrate-specific electron-phonon coupling constant and $D$, the average distance between the substrate and the material under consideration. $D$ will be different depending on whether we are considering SL graphene, BL graphene or a semi-Dirac. For SL and BL graphene $D$ are about $4 \AA$ and $6 \AA$ respectively\cite{PolGAndBLG}. For semi-Dirac, an average value of $D$ can be taken to be about $15 \AA$ for the following reason. Pardo and Pickett discovered, while studying $(TiO_2)_5/(VO_2)_3$ hetero-structure\cite{PardoSd}, that the semi-Dirac band-structure sports the signature of the Vanadium atoms, which in the real space are located above $5$ layers of $TiO_2$ of about a total of $1.5$nm thickness. Hence, if  $(TiO_2)_5/(VO_2)_3$ layered structure is grown on a polar substrate, the semi-Dirac electrons will be separated from the substrate by at least $1.5$nm thick $TiO_2$ layers. It is noted that for small $q$, ${M_{\vec{q}}}^2$ in the above goes as $\frac{1}{q}$, just like an electrostatic potential in two dimension. 
\\
\\
Finally, using Eq.~\ref{eq:ElecDens} along with the above-mentioned expression for $M_{\vec{q}}$, in Eq.~\ref{eq:PolaronHamil} one obtains 
\begin{equation}
\begin{aligned}
H = \sum_{\vec{q}, \vec{k}} \sqrt{g}\frac{e^{-qD}}{\sqrt{q}} \Big[ u^{\dagger}_{\vec{k}}u_{\vec{k}+\vec{q}}  c^{\dagger}_{\vec{k}}c_{\vec{k}+\vec{q}}{b^{\dagger}_{\vec{q}}}  + h.c \Big ]
\label{eq:PolaronHamilFinal}
\end{aligned}
\end{equation}
Eq.~\ref{eq:PolaronHamilFinal} is the long-wavelength dominated electron-phonon interaction Hamiltonian that will be used for all our subsequent calculations for all the three systems. It is seen from Eq.~\ref{eq:PolaronHamilFinal} that this Hamiltonian has both the Fermionic and the Bosonic operators multiplying each other, which represents the electron-phonon interaction. It is noted that the eigen-spinors in Eq.~\ref{eq:PolaronHamilFinal} correspond to the positive energy of the `non-interacting' Hamiltonian. \footnotemark[1] \footnotetext[1]{The negative energy eigen-spinors and the associated hole-operators play no role here. This has to do with the fact that the polaron formation involves the initial and the final state being an electron state. It can be shown while calculating the polaron energy from Eq.~\ref{eq:EnergyCorrectionOne} that the presence of any hole operators in H, and/or any hole state for the intermediate state $\ket {n}$ will not lead to a finite non-zero result, given that only an electron of a certain momentum is in the initial as well as in the final state.} Using the language of the perturbation theory involving second quantization, the polaron formation can be explained with the help of Eq.~\ref{eq:PolaronHamilFinal} in the following way.
We will destroy an existing electron, and while doing so, create an electron-phonon pair as per the Hamiltonian given by Eq.~\ref{eq:PolaronHamilFinal}, and then subsequently destroy that electron-phonon pair and create an electron having the same momentum as the original electron, once again following the same Hamiltonian.  
\\
\\
Finally, the constant $g$ appearing in Eq.~\ref{eq:PolaronHamilFinal} can be investigated further. Following Wang and Mahan\cite{Mahan}, $g$ can be written as follows. 
\begin{equation}
\begin{aligned}
g = \frac{2\pi \tilde{\epsilon} e^2 \hbar \omega_s}{A}, 
\label{eq:ExpressionG}
\end{aligned}
\end{equation}
 
where $\tilde{\epsilon} = \frac{\epsilon_s - \epsilon_{\infty}}{(\epsilon_s+1)(\epsilon_{\infty}+1)}$, $\epsilon_s$ and $\epsilon_{\infty}$ being the static and high frequency permittivities respectively of the substrate. $\omega_s$ is the frequency of the optical phonons at the surface of the polar substrate; $e$ is the charge of an electron; and $A$ is the physical area of the system. Quantities like $\tilde{\epsilon}$ and $\omega_s$ are substrate specific. As an example, for $6H-SiC$ substrate, $\omega_s = 116 meV$\cite{PolGAndBLG}. Also, for $6H-SiC$, $\epsilon_s = 9.7$, and $\epsilon_{\infty} = 6.5$\cite{PolGAndBLG}, which gives $\tilde{\epsilon} = .04$ for $6H-SiC$. For other substrates there will be other values for the above mentioned parameters. A detailed discussion on the topic is given in [\onlinecite{PolGAndBLG}].

 \section{Deriving the general polaron energy expression (for SL/BL graphene and semi-Dirac)}
Next we will obtain an expression for the energy correction for an electron of wave-vector $\vec{l}$, due to the interaction Hamiltonian given by Eq.~\ref{eq:PolaronHamilFinal}, the initial and the final wave-vector $\vec{l}$ of the electron remaining unchanged. Since there are no phonons in the initial and the final state, and the interaction Hamiltonian has either the phonon creation or the annihilation operator, but not both in each of its terms, the first order energy correction will be zero. This can clearly be seen from the fact that $\bra {0\text{  phonon}} b_{\vec{q}} (\text{or } b^{\dagger}_{\vec{q}} ) \ket{ 0\text{  phonon}} = 0$. The first non-vanishing electron energy correction will arise from the second order perturbation expression for the energy. 

\begin{tikzpicture}
\begin{feynman}
\vertex (a);
\vertex[right=1.5cm of a] (b);
\vertex[right=2.5cm of b] (e);
\vertex[right=1.5cm of e] (f);
\diagram*{(a)--[plain] (f),
(a) --[momentum'= \(l\)] (b),
(b) -- [photon, momentum' = \(q\), half left] (e), 
(b) --[momentum'= \(l-q\)] (e),
(e) --[momentum'= \(l\)] (f)
};
\end{feynman}
\end{tikzpicture}

The above figure is a Feynman diagram showing a second-order electron-phonon interaction, in which a phonon, shown as a wavy line in the diagram, is emitted and then re-absorbed. The solid lines in the diagram represent electrons. In the Feynman diagram, despite the interaction, the electron's initial and the final states are the same having the same wave-vector $\vec{l}$. At each node of the Feynman diagram, the momentum conservation law is obeyed as indicated in the diagram. The above Feynman diagram serves as a visual aid, not used for the calculation of the change in the energy of the electron, or the self energy, due to such a process. The self energy ($\Delta E$) of the electron is calculated using the second order perturbation energy expansion expression involving the second-quantized Hamiltonian operator given by Eq.~\ref{eq:PolaronHamilFinal}. The expression for $\Delta E$ is as follows:

\begin{equation}
\begin{aligned}
&\Delta E = \\
& -\sum_{n} \frac{\bra{\text{$\vec{l}^{el}$, no phonons}} H \ket {n} \bra{n} H \ket {\text{$\vec{l}^{el}$, no phonons}}}{E_n - E_0} 
\label{eq:EnergyCorrectionOne}                              
\end{aligned}
\end{equation}
 In Eq.~\ref{eq:EnergyCorrectionOne}, $\ket {\text{$\vec{l}^{el}$, no phonons}}$ is the initial as well as the final state, which has only one electron of wave-vector $\vec{l}$ and no phonons. This state in the second quantized notation can be written as, $\ket {\text{$\vec{l}^{el}$, no phonons}} = c_{\vec{l}}^{\dagger}\ket{0}$, where $\ket{0}$ corresponds to the `vacuum'. [`vacuum' refers to the absence of the electron of interest, as well as the absence of any phonons]. Denoting the energy of the sole electron that the initial (as well as the final) state comprises of as $E^{el}_{\vec{l}}$, $E_0$ in Eq.~\ref{eq:EnergyCorrectionOne} can be written as $E_0 = E^{el}_{\vec{l}}$.  
\\
\\
$\ket{n}$ stands for the intermediate state which can be written as $\ket{n} \equiv \ket{\text{$\vec{p_1}^{ph}$, $\vec{p_2}^{el}$}}$, where $\vec{p_1}^{ph}$ stands for the wave-vector of the phonon, and $\vec{p_2}^{el}$ stands for the wave-vector of the electron. In the second quantized notation, the  intermediate state $\ket{n}$ can be written as $\ket{n} = {b^{\dagger}_{\vec{p_1}}} {c^{\dagger}_{\vec{p_2}}} \ket{0}$, $b^{\dagger}_{\vec{p_1}}$  and $c^{\dagger}_{\vec{p_2}}$ corresponding to the phonon and electron creation operators respectively. $\sum_{n}$ in Eq.~\ref{eq:EnergyCorrectionOne} is actually $\sum_{\vec{p_1}, \vec{p_2}}$.
\\
\\
$E_n$, the energy of the intermediate state $\ket{n}$, can be written as the sum of the energy of the `free' electron (denoted as $E^{el}_{\vec{p_2}}$) and that of the phonon with wave-vector $\vec{p_1}$. The phonon being an optical phonon, has a momentum-independent constant frequency $\omega_s$, and energy $\hbar \omega_s$. Hence $E_n$ can be written as $E_n = E^{el}_{\vec{p_2}} + \hbar \omega_s$. We obtain from Eq.~\ref{eq:EnergyCorrectionOne} after having put all the pieces together, the following expression for the self energy. 

\begin{equation}
\begin{aligned}
&\Delta E = \\
& -\sum_{\vec{p_1}, \vec{p_2}, \vec{q}, \vec{k}} g\frac{e^{-2qD}}{q}
\frac{| \bra{0}c_{\vec{p_2}} b_{\vec{p_1}}
[u^{\dagger}_{\vec{k}}u_{\vec{k}+\vec{q}}  c^{\dagger}_{\vec{k}}c_{\vec{k}+\vec{q}}{b^\dagger_{\vec{q}}}] c^{\dagger}_{\vec{l}}\ket{0} |^2}
{E^{el}_{\vec{p_2}} +\hbar \omega_s - E^{el}_{\vec{l}} } 
\label{eq:EnergyCorrectionTwo}
\end{aligned}
\end{equation}
\\
\\
Next we will use the standard commutation relations for Bosonic $b$ operators and anticommutation relations for the Fermionic $c$ operators respectively in equation Eq.~\ref{eq:EnergyCorrectionTwo}, such as $[b_{\vec{p_1}}, {b^{\dagger}_{\vec{q}}}] = \delta_{\vec{p_1}, \vec{q}}$, $\{c_{\vec{k}}, c_{\vec{l}}^{\dagger} \} = \delta_{\vec{k}, \vec{l}}$, $[b_{\vec{p_1}}, c_{\vec{l}}^{\dagger}] = 0$ etc. Thus we obtain the following expression for the self energy, or the polaron energy as it will be referred to.
\begin{equation}
\begin{aligned}
& \Delta E = -\sum_{\vec{q}} g\frac{e^{-2qD}}{q}
\frac{| u^{\dagger}_{\vec{l}-\vec{q}}u_{\vec{l}} |^2}
{E^{el}_{\vec{l}-\vec{q}} +\hbar \omega_s - E^{el}_{\vec{l}}} 
\label{eq:EnergyCorrectionFinal}
\end{aligned}
\end{equation}
The factor $| u^{\dagger}_{\vec{l}-\vec{q}}u_{\vec{l}} |^2$ in the numerator of Eq.~\ref{eq:EnergyCorrectionFinal} describes the overlap of the spinors, also called the overlap factor. The spin degeneracy factor $g_s$ needs to be multiplied to Eq.~\ref{eq:EnergyCorrectionFinal} to make the energy expression complete. We will insert it in the final energy expressions for the individual materials.

\section{Analytical derivation of the polaron energy for the case of BL-graphene}
In this section we will compute the polaron energy as given by Eq.~\ref{eq:EnergyCorrectionFinal} for a BL graphene-electron. The `non-interacting' Hamiltonian of BL graphene in the momentum space $(\hbar p_x, \hbar p_y)$ is given by: \\
$H^{0}_{BLG}\equiv \frac{\hbar^2}{2m}\begin{pmatrix}
0 &  p_{x}^2-p_{y}^2-i2p_xp_y \\
p_{x}^2-p_{y}^2+i2p_xp_y & 0
\end{pmatrix}$, 
with the eigen-energies $E^{el}_{\vec{p}} = \pm \frac{\hbar^2 p^2}{2m}$, the $+$ sign corresponding to the electron-eigenspinor 
\begin{eqnarray}
u_{\vec{p}} =\frac{1}{\sqrt{2}} \begin{pmatrix}
1 \\
e^{i2\theta_{\vec{p}}} 
\end{pmatrix},
\label{eq:BilayerGSpinor}
\end{eqnarray}.

where $\theta_{\vec{p}} = \tan^{-1} \frac{p_y}{p_x}$. The overlap factor in Eq.~\ref{eq:EnergyCorrectionFinal} can be calculated for the BL graphene using the spinors given by Eq.~\ref{eq:BilayerGSpinor}. To that end, the overlap factor is written as follows.  
\begin{equation}
\begin{aligned}
| u^{\dagger}_{\vec{l}-\vec{q}}u_{\vec{l}}|^2 = u^{\dagger}_{\vec{l}}u_{\vec{l}-\vec{q}}u^{\dagger}_{\vec{l}-\vec{q}}u_{\vec{l}} 
\label{eq:OverlapBLG}
\end{aligned}
\end{equation}
Next with an aim to using it in  Eq.~\ref{eq:OverlapBLG}, the following identity involving the BL graphene spinors is established using Eq.~\ref{eq:BilayerGSpinor}.
\begin{equation} 
\begin{aligned}
& u_{\vec{p}}u^{\dagger}_{\vec{p}} = \frac{1}{2}\left [I + \frac{1}{|\vec{p}|^2}\{ ({p_x}^2-{p_y}^2)\sigma_x + 2 p_x p_y\sigma_y\}\right ]
\label{eq:BLGSpinorIdentity}
\end{aligned}
\end{equation}
($\sigma$s are the Pauli matrices and $I$, the 2 by 2 identity matrix). Using  Eq.~\ref{eq:BLGSpinorIdentity} with $\vec{p} = \vec{l}-\vec{q}$, in Eq.~\ref{eq:OverlapBLG}  one obtains 
\begin{equation}
\begin{aligned}
 |u^{\dagger}_{\vec{l}-\vec{q}}u_{\vec{l}} |^2 &= \\
 u^{\dagger}_{\vec{l}} \frac{1}{2}  [I & + \frac{1}{|\vec{l}-\vec{q}|^2}\{ ({(l_x-q_x)}^2-{(l_y-q_y)}^2)\sigma_x \\
 & +  2 (l_x-q_x) (l_y-q_y)\sigma_y\} ] u_{\vec{l}}
 \label{eq:OverlapBLGTwo}
\end{aligned}
\end{equation}
Instead of continuing to use the cartesian components, e.g., $l_x, l_y, q_x$ and $q_y$ in Eq.~\ref{eq:OverlapBLGTwo}, we will express $\vec{l}$ and $\vec{q}$ in terms of polar co-ordinates in order to facilitate subsequent calculations. The electron wave-vector $\vec{l}$ of magnitude $|\vec{l}| \equiv l$ is assumed to make an angle $\gamma$ with the $x$ direction. Also, an angle $\phi$ is defined between the wave-vector $\vec{q}$ of magnitude $|\vec{q}| \equiv q$ and the $x$ direction. The quantities $l_x, l_y, q_x$ and $q_y$, which show up in Eq.~\ref{eq:OverlapBLGTwo}, can be rewritten in terms of $l$, $q$, and angular variables $\gamma$ and $\phi$ as follows. 

\begin{equation}
\begin{aligned}
l_x = l\cos \gamma, l_y = l\sin \gamma, q_x = q\cos \phi, q_y = q\sin \phi 
\label{eq:LQGammaPhi}
\end{aligned}
\end{equation}
With the electron wavevector $\vec{l}$ making an angle $\gamma$ with the $x$ axis, the spinor $u_{\vec{l}}$ will be given, as per  Eq.~\ref{eq:BilayerGSpinor}, by 
\begin{equation}
\begin{aligned}
&u_{\vec{l}} = \frac{1}{\sqrt{2}}\begin{pmatrix}
& 1 \\
& e^{i2\gamma}
\end{pmatrix}
\label{eq:BLGSpinorGamma}
\end{aligned}
\end{equation} 

Using Eq.~\ref{eq:LQGammaPhi} and Eq.~\ref{eq:BLGSpinorGamma} in Eq.~\ref{eq:OverlapBLGTwo}, after some algebra, one obtains 
\begin{equation}
\begin{aligned}
 |u^{\dagger}_{\vec{l}-\vec{q}}u_{\vec{l}} |^2 =\Bigg[1-\frac{q^2\sin^2(\phi-\gamma)}{|\vec{l}-\vec{q}|^2}\Bigg]
 \label{eq:OverlapBLGFinal}
\end{aligned}
\end{equation}
Finally, using BL graphene electron energies $E^{el}_{\vec{l}-\vec{q}} = \frac{\hbar^2|\vec{l}-\vec{q}|^2}{2m}$, and $E^{el}_{\vec{l}} =\frac{ \hbar^2l^2}{2m}$ in Eq.~\ref{eq:EnergyCorrectionFinal}, as well as using Eq.~\ref{eq:OverlapBLGFinal} in Eq.~\ref{eq:EnergyCorrectionFinal}, one obtains the following expression for the BL graphene polaron energy $(\equiv \Delta E_{\text{BLG}})$.

\begin{equation}
\begin{aligned}
& \Delta E_{\text{BLG}} = -\sum_{\vec{q}} g\frac{e^{-2qD}}{q}
\frac{\Bigg[1-\frac{q^2\sin^2(\phi-\gamma)}{|\vec{l}-\vec{q}|^2}\Bigg]}
{ \frac{{\hbar^2\abs{\vec{l}-\vec{q}}}^2}{2m}+\hbar\omega_s - \frac{\hbar^2l^2}{2m}} 
\label{eq:EnergyBLGCorrectionOne}
\end{aligned}
\end{equation}
In Eq.~\ref{eq:EnergyBLGCorrectionOne}, the angular variable $(\phi-\gamma)$ shows up explicitly in the expression $\sin^2(\phi-\gamma)$. $(\phi-\gamma)$ is also present in the expression $|\vec{l}-\vec{q}|^2$ in the same equation. That can clearly be seen by writing out $|\vec{l}-\vec{q}|^2$ as $l^2 + q^2 -2lq\cos(\phi - \gamma)$. $(\phi-\gamma)$ is the angle between the vectors $\vec{l}$ and $\vec{q}$. A new angular variable $\phi^\prime \equiv (\phi-\gamma)$ can now be introduced and  $\sum_{\vec{q}}$ can be replaced as $A\int_0^\infty qdq \int_0^{2\pi} \frac{d\phi^\prime}{(2\pi)^2}$, the integrand of the double integral depending on the variable $q$, and $\phi^\prime$. Doing so, Eq.~\ref{eq:EnergyBLGCorrectionOne} assumes the following form.

\begin{equation}
\begin{aligned}
&\Delta E_{\text{BLG}}\\
& = -gA \int_0^\infty dq \int_0^{2\pi} d\phi^\prime e^{-2qD} 
\frac{\Bigg[1-\frac{q^2\sin^2 \phi^\prime}{|\vec{l}-\vec{q}|^2}\Bigg]}
{ \frac{{\hbar^2\abs{\vec{l}-\vec{q}}}^2}{2m}+\hbar\omega_s - \frac{\hbar^2l^2}{2m}} 
\label{eq:EnergyBLGPhiInt}
\end{aligned}
\end{equation}
\\
\\
At this point we define the following dimensionless quantities: $\vec{l^\prime} = \frac{\vec{l}}{\sqrt{\frac{m\omega_s}{\hbar}}}$  and $\vec{q^\prime} = \frac{\vec{q}}{\sqrt{\frac{m\omega_s}{\hbar}}}$. We call them `dimensionless electron momenta' or `DEM's, since they are proportional to the momenta variables $\hbar \vec{l}$ and $\hbar \vec{q}$ respectively. We replace $\vec{l}$ and $\vec{q}$ in Eq.~\ref{eq:EnergyBLGPhiInt} by these dimensionless variables. Finally, using the expression for $g$ as given by Eq.~\ref{eq:ExpressionG} in Eq.~\ref{eq:EnergyBLGPhiInt}, we obtain the following equation for the polaron energy for BL graphene.

\begin{equation}
\begin{aligned}
&\Delta E_{\text{BLG}} =-\frac{\tilde{\epsilon}e^2}{\pi}\sqrt{\frac{m\omega_s}{\hbar}}(I_1 - I_2),
\label{eq:EnergyBLGDimlessShort}
\end{aligned}
\end{equation}

where 
\begin{subequations}
\begin{equation}
\begin{aligned}
&I_1 = \int_0^\infty dq^\prime e^{-2q^\prime \tilde d} \int_0^{2\pi} d\phi^\prime \frac{1}{|\vec{l^\prime}-\vec{q^\prime}|^2+2-|\vec{l^\prime}|^2} \\
\label{eq:I_One}
\end{aligned}
\end{equation}

\begin{equation}
\begin{aligned}
&I_2 = \int_0^\infty dq^\prime e^{-2q^\prime \tilde d} \int_0^{2\pi} d\phi^\prime \frac{{q^\prime}^2 \sin^2 \phi^\prime}{{|\vec{l^\prime}}-\vec{q^\prime}|^2[|\vec{l^\prime}-\vec{q^\prime}|^2+2-|\vec{l^\prime}|^2]}
\label{eq:I_Two}
\end{aligned}
\end{equation}
\label{eq:I_1And I_2}
\end{subequations}
\\
\\
$\tilde d$ in the above expressions is a dimensionless quantity given by $\tilde d = \sqrt{\frac{m\omega_s}{\hbar}}D$. We will consider the specific case of BL graphene grown on $6H-SiC$ substrate, for which $\omega_s = 116 meV$\cite{PolGAndBLG}. $D=6\AA$ for BL graphene\cite{PolGAndBLG}. These values along with the standard values of electron-mass $m$ and $\hbar$ yield $\tilde d \approx .74$. Other values of $\tilde d$ can be obtained for other substrates, but as long as they are of the same order of magnitude, the essential physics will change little. For all our calculations involving BL graphene we will use $\tilde d = .74$. 
\\
\\
In order to find $\Delta E_{\text{BLG}}$ in Eq.~\ref{eq:EnergyBLGDimlessShort}, we would need to evaluate the integrals given by $I_1$ and $I_2$. We will first do so analytically for small dimensionless electron momentum (DEM) $l^\prime$. As will be shown in the following, the small $l^\prime$ expansion of $\Delta E_{\text{BLG}}$ will not only have a constant and a term quadratic in $l^\prime$, but it will also have a term linear in  $l^\prime$. 
\\
\\
In both $I_1$ and $I_2$, the factor $\frac{1}{|\vec{l^\prime}-\vec{q^\prime}|^2+2-|\vec{l^\prime}|^2}$ appears. It can be simplified to give $\frac{1}{2+{q^\prime}^2-2l^\prime q^\prime \cos \phi^\prime}$. Next following Kiittel{\cite{Kittel}}, since $l^\prime \ll 1$, it is possible to write $(2+{q^\prime}^2) \gg 2l^\prime q^\prime \cos \phi^\prime$, which is true for all $q^\prime$ and $\phi^\prime$. Hence, $\frac{1}{2+{q^\prime}^2-2lq\cos \phi^\prime}$ can be Taylor-expanded as follows. 
\begin{equation}
\begin{aligned}
&\frac{1}{2+{q^\prime}^2-2l^\prime q^\prime \cos(\phi^\prime)} \\
& = \frac{1}{(2+{q^\prime}^2)[1-\frac{2l^\prime q^\prime \cos(\phi^\prime)}{2+{q^\prime}^2}]} \\
& \approx \frac{1}{2+{q^\prime}^2} \Big [1 + \frac{2l^\prime q^\prime \cos(\phi^\prime)}{2+{q^\prime}^2} + \frac{4{l^\prime}^2{q^\prime}^2\cos^2(\phi^\prime)}{(2+{q^\prime}^2)^2}+...\Big]
\label{eq:TaylorExpansion}
\end{aligned}
\end{equation}   
We will use the series expansion as given by Eq.~\ref{eq:TaylorExpansion}, to calculate $I_1$ and $I_2$ in Eq.~\ref{eq:I_1And I_2}. We focus on the calculation of $I_1$ first. Using Eq.~\ref{eq:TaylorExpansion} in  Eq.~\ref{eq:I_One}, and keeping terms up to $o({l^\prime}^2)$, we obtain the following.   
\begin{equation}
\begin{aligned}
I_1 & \approx 
 \int_0^\infty dq^\prime \int_0^{2\pi} d\phi \frac{e^{-2q^\prime \tilde d}}{({q^\prime}^2+2)}\Bigg[1+\frac{2l^\prime q^\prime \cos \phi^\prime}{{q^\prime}^2+2} \\
& \hspace{55 mm} +\frac{4{l^\prime}^2{q^\prime}^2\cos^2 \phi ^\prime}{({q^\prime}^2+2)^2}\Bigg]
\label{eq:I_OneQPhi}
\end{aligned}
\end{equation}
Carrying out the $\phi{^\prime}$ integrals in Eq.~\ref{eq:I_OneQPhi}, which is straightforward, one obtains the following expression for $I_1$
\begin{equation}
\begin{aligned}
& I_1 \approx 2\pi \int_0^\infty dq^\prime \frac{e^{-2q^\prime \tilde{d}}}{{q^\prime}^2+2}  + 4\pi {l^\prime}^2\int_0^\infty dq^\prime \frac{e^{-2q^\prime \tilde{d}}{q^\prime}^2}{({q^\prime}^2+2)^3}
\label{eq:I_1FinalForm}
\end{aligned}
\end{equation}
\\
\\
Next we focus on evaluating $I_2$. Using the Taylor expansion, as given by Eq.~\ref{eq:TaylorExpansion}, in Eq.~\ref{eq:I_Two} we obtain the following series-expansion expression for $I_2$
\begin{equation}
\begin{aligned}
I_2 = J_1 + J_2 + J_3.......   ,
\label{eq:I_2Expansion}
\end{aligned}
\end{equation}
where 

\begin{subequations}
\begin{equation}
J_1 = \int_0^\infty dq^\prime e^{-2q^\prime \tilde{d}} \frac{{q^\prime}^2}{{q^\prime}^2+2} \int_0^{2\pi} d\phi^\prime \frac{\sin^2 \phi^\prime}{|\vec{l^\prime}-\vec{q^\prime}|^2} 
\label{eq:JOneQPhi}
\end{equation}
\begin{equation}
J_2 
= \int_0^\infty dq^\prime e^{-2q^\prime \tilde{d}}\frac{2{l^\prime q^\prime}^3}{({q^\prime}^2+2)^2}\int_0^{2\pi} d\phi^\prime \frac{\sin^2 \phi^\prime \cos \phi^\prime }{|\vec{l^\prime}-\vec{q^\prime}|^2}
\label{eq:JTwoQPhi}
\end{equation}
\begin{equation}
 J_3 
= \int_0^\infty dq^\prime e^{-2q^\prime \tilde{d}}\frac{4{l^\prime}^2 {q^\prime}^4}{({q^\prime}^2+2)^3}\int_0^{2\pi} d\phi^\prime  \frac{\sin^2 \phi^\prime \cos^2 \phi^\prime}{|\vec{l^\prime}-\vec{q^\prime}|^2}
\label{eq:JThreeQPhi}
\end{equation}
\label{eq:JOneTwo}
\end{subequations}
In Eq.~\ref{eq:JOneTwo}, we kept only the terms which contributes up to $o({l^\prime}^2)$ to $\Delta E_{\text{BLG}}$.  Computing $J_1$, $J_2$ and $J_3$ analytically for small $l^\prime$ is rather involved, and hence is discussed in Appendix~\ref{app:app1}. In the following we simply mention the final result (upto $o({l^\prime}^2)$) for $J_1$, $J_2$, and $J_3$, 

\begin{equation}
\begin{aligned}
& J_1 \approx  \pi \int_0^\infty dq^\prime \frac{e^{-2q^\prime \tilde{d}}}{{q^\prime}^2+2} -\frac{\pi}{3}l^\prime + \frac{\pi \tilde{d}}{4} {l^\prime}^2 \\
& J_2 \approx \pi{l^\prime}^2\int_0^\infty dq^\prime \frac{e^{-2q^\prime \tilde{d}}}{({q^\prime}^2+2)^2}\\
& J_3 \approx \pi {l^\prime}^2 \int_0^\infty dq^\prime \frac{e^{-2q^\prime \tilde{d}}{q^\prime}^2}{({q^\prime}^2+2)^3}
\label{eq:JOneTwoFinal}
\end{aligned}
\end{equation}

\begin{figure}[ht]
\begin{center}
\includegraphics[draft=false,bb=0 0 550 550, clip, width=\columnwidth]{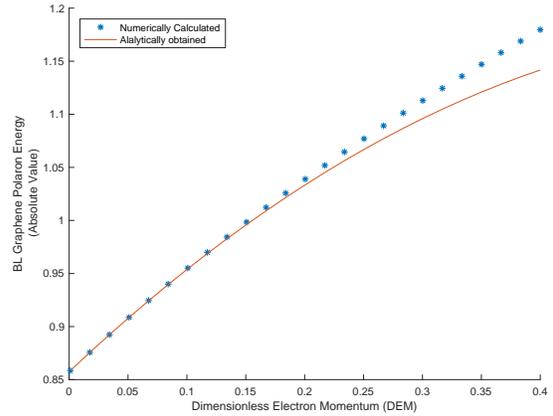}
\caption{Absolute Value of Polaron Energy (in the units of $\frac{g_s\tilde{\epsilon}e^2}{\pi}\sqrt{\frac{m\omega_s}{\hbar}}$) Versus Dimensionless Electron Momentum (DEM) $l^\prime$ for BL graphene for small $l^\prime$.}
\label{fig:BLGEnergy}
\end{center}
\end{figure}

Using Eq.~\ref{eq:JOneTwoFinal}, Eq.~\ref{eq:I_2Expansion} and Eq.~\ref{eq:I_1FinalForm} in  Eq.~\ref{eq:EnergyBLGDimlessShort} we obtain the following final expression for $\Delta E_{\text{BLG}}$.
\begin{equation}
\begin{aligned}
&\Delta E_{\text{BLG}} \\
&\approx -\frac{\tilde{\epsilon}e^2}{\pi}\sqrt{\frac{m\omega_s}{\hbar}} 
\Bigg[\pi \int_0^\infty dq^\prime \frac{e^{-2q^\prime \tilde{d}}}{{q^\prime}^2+2} +\frac{\pi}{3}l^\prime \\
& +{l^\prime}^2\Bigg(3\pi \int_0^\infty dq^\prime \frac{e^{-2q^\prime \tilde{d}}{q^\prime}^2}{{(q^\prime}^2+2)^3}
-\pi \int_0^\infty dq^\prime \frac{e^{-2q^\prime \tilde{d}}}{({q^\prime}^2+2)^2} \\
&\hspace{55mm}-\frac{\pi \tilde{d}}{4} \Bigg) \Bigg] 
 \label{eq:EnergyFinalNoApprox}
\end{aligned}
\end{equation}
The integrals appearing in Eq.~\ref{eq:EnergyFinalNoApprox} can be readily evaluated numerically. Using the value of $\tilde{d} = .74$ one obtains $\int_0^\infty dq^\prime \frac{e^{-2q^\prime \tilde{d}}}{{q^\prime}^2+2} = .2729, \int_0^\infty dq^\prime \frac{e^{-2q^\prime \tilde{d}}{q^\prime}^2}{{(q^\prime}^2+2)^3} = .0118$, and $\int_0^\infty dq^\prime \frac{e^{-2q^\prime \tilde{d}}}{{(q^\prime}^2+2)^2} = .1184$. Using these integrals in Eq.~\ref{eq:EnergyFinalNoApprox}, and inserting the spin degeneracy factor $g_s$, we obtain
\begin{equation}
\begin{aligned}
&\Delta E_{\text{BLG}} \approx \\
&-\frac{g_s\tilde{\epsilon}e^2}{\pi}\sqrt{\frac{m\omega_s}{\hbar}}\Bigg[.2729\pi+\frac{\pi}{3}l^\prime -(.2679 \pi){l^\prime}^2 \Bigg]
\label{eq:EnergyFinal}
\end{aligned}
\end{equation}
\\
\\
In Eq.~\ref{eq:EnergyFinal} there is a constant term, which corresponds to the polaron energy when the DEM $l^\prime$ approaches 0. Additionally, in Eq.~\ref{eq:EnergyFinal} there is a term linear in the DEM and there is a term quadratic in the DEM. For small DEM it's the linear term that dominates. In other words, BL graphene polaron energy disperses linearly with electron momenta, for small values of the latter. This makes BL graphene polaron massless for small DEM. This is the key result of this section.
\\
\\
It can be seen from Eq.~\ref{eq:EnergyBLGPhiInt} that the polaron energy for BL graphene becomes singular when the denominator  $\frac{\hbar^2|\vec{l}-\vec{q}|^2}{2m}+\hbar\omega_s-\frac{\hbar^2|\vec{l}|^2}{2m}$ is $0$. This, in terms of the DEMs $l^\prime$ and $q^\prime$ as introduced before, simply becomes $|\vec{l^\prime}-\vec{q^\prime}|^2+2-|\vec{l^\prime}|^2 = 0$. With a little algebra, it can be shown that for $l^\prime < \sqrt{2}$, $|\vec{l^\prime}-\vec{q^\prime}|^2+2-|\vec{l^\prime}|^2 $ is always greater than $0$ (and hence not $0$) for any $q^\prime$ and $\phi^\prime$. Hence, there is no singularity in the polaron energy given by Eq.~\ref{eq:EnergyBLGPhiInt}, when $l^\prime$ is less than $\sqrt{2}$. In other words, polaron-formation in the context of BL graphene is theoretically guaranteed as long as the DEM $l^\prime$ is less than the cut-off value $\sqrt{2}$.  
\\
\\
Incidentally, one might be tempted to think that in Eq.~\ref{eq:EnergyBLGPhiInt} there is another possibility of singularity coming from the factor $\frac{1}{|\vec{l}-\vec{q}|^2}$ in the numerator of the right side of the equation, becoming infinite when $\vec{l}-\vec{q} = 0$, or $\vec{l}=\vec{q}$. But on close inspection, it is seen that $\vec{l}=\vec{q}$ is not a true singularity in the sense that the factor $\sin^2\phi^\prime$ multiplying $\frac{1}{|\vec{l}-\vec{q}|^2}$ is zero, when $\vec{l}=\vec{q}$. [$\phi^\prime$ is $0$, when $\vec{q}$ is in the same direction as $\vec{l}$]
\\
\\
In Fig.~\ref{fig:BLGEnergy} the absolute value of the polaron energy $\Delta E_{\text{BLG}}$ given by Eq.~\ref{eq:EnergyFinal} is plotted w.r.t $l^\prime$. In the same figure the absolute value of $\Delta E_{\text{BLG}}$, evaluated numerically by using Eq.~\ref{eq:EnergyBLGDimlessShort} directly, is also plotted w.r.t $l^\prime$ for comparison. The energy axis in the plot is in the units of $\frac{g_s\tilde{\epsilon}e^2}{\pi}\sqrt{\frac{m\omega_s}{\hbar}}$.
It can be seen from Fig.~\ref{fig:BLGEnergy} that the analytical result agrees with the numerical result as long as $l^\prime$ isn't too large.

\section{General Expression for the Decay Rate} 
In the last section we argued that the polaron formation in the context of BL graphene happens when the DEM $l^\prime < \sqrt{2}$, since then the denominator in Eq.~\ref{eq:EnergyBLGPhiInt} is guaranteed to not become zero. This is no longer the case when $l^\prime > \sqrt{2}$. The vanishing of the denominator in Eq.~\ref{eq:EnergyBLGPhiInt} is then a real possibility, thereby making the BL Graphene polaron energy expression undefined. In this section we will address the vanishing of the denominator in the general polaron energy expression given by Eq.~\ref{eq:EnergyCorrectionOne}, of which Eq.~\ref{eq:EnergyBLGPhiInt} is a special case. This treatment will be applicable for SL graphene and semi-Dirac as well. Following the standard procedure, the issue of the vanishing denominator is taken care of by analytically continuing the denominator to the complex plane\cite{Feynman} as shown in the following. 

\begin{equation}
\begin{aligned}
\frac{1}{E_n-E_0-i \epsilon} = PV \Bigg[\frac{1}{E_n-E_0} \Bigg] +i\pi \delta (E_n-E_0)
\label{eq:BLGEnergyNDelta}
\end{aligned}
\end{equation}
In Eq.~\ref{eq:BLGEnergyNDelta}, $PV$ is the principal value, which is defined when $E_n \neq E_0$. The imaginary part of Eq.~\ref{eq:BLGEnergyNDelta} involves a delta function and is non-zero only when $E_n = E_0$. Eq.~\ref{eq:BLGEnergyNDelta}, when inserted in Eq.~\ref{eq:EnergyCorrectionOne}, produces an imaginary component in the expression for the polaron energy, which would ultimately give us the decay rate. The imaginary component of the polaron energy(defined as $\Delta E_{I}$) is obtained as

\begin{equation}             
\begin{aligned}
&\Delta E_{I} \\
& = -i\pi\sum_{n} |\bra{n} H \ket {\text{$\vec{l}_{el}$, no phonons}}|^2 \delta (E_n-E_0)
\label{eq:ImagEnergy}
\end{aligned}
\end{equation}
Following steps similar to those taken while obtaining Eq.~\ref{eq:EnergyCorrectionFinal} from Eq.~\ref{eq:EnergyCorrectionOne}, $\sum_{n}$ in Eq.~\ref{eq:ImagEnergy} is replaced by $\sum_{\vec{q}}$, $|\bra{n} H \ket {\text{$\vec{l}_{el}$, no phonons}}|^2$ is replaced by $g\frac{e^{-2qD}}{q}| u^{\dagger}_{\vec{l}-\vec{q}}u_{\vec{l}} |^2$, and $(E_n-E_0)$ is replaced by $(E^{el}_{\vec{l}-\vec{q}} +\hbar \omega_s - E^{el}_{\vec{l}})$. Thus, Eq.~\ref{eq:ImagEnergy} assumes the following form.
\begin{equation}             
\begin{aligned}
&\Delta E_{I} \\
& = -i\pi\sum_{\vec{q}} g\frac{e^{-2qD}}{q}
| u^{\dagger}_{\vec{l}-\vec{q}}u_{\vec{l}} |^2
\delta (E^{el}_{\vec{l}-\vec{q}} +\hbar \omega_s - E^{el}_{\vec{l}})
\label{eq:ImagEnergyFinal}
\end{aligned}
\end{equation}
The decay rate $(\equiv \frac{1}{\tau})$ is $\frac{2|\Delta E_I|}{\hbar}$, which can be seen as follows. Writing $\Delta E_I=-i|\Delta E_I|$ in the standard quantum mechanical expression $e^{-\frac{i\Delta Et}{\hbar}}$, yields an exponentially decaying function $e^{-\frac{|\Delta E_I| t}{\hbar}}$ due to $\Delta E_I$ being imaginary. The exponentially decaying function corresponds to the probability of the electron leaving its existing state, its energy dissipating due to the creation of a phonon. More specifically, the transition probability ($\equiv P$) is obtained by squaring $e^{-\frac{|\Delta E_I| t}{\hbar}}$. This gives $P \equiv e^{-\frac{2|\Delta E_I| t}{\hbar}}$, from which the decay rate $\frac{1}{\tau}$ can be read off as $\frac{2}{\hbar}|\Delta E_I|$.  From Eq.~\ref{eq:ImagEnergyFinal} after replacing $\sum_{\vec{q}}$ by $A\frac{d^2\vec{q}}{(2\pi)^2}$, one obtains the following general expression for the decay rate $\frac{1}{\tau}$. 

\begin{equation}             
\begin{aligned}
\frac{1}{\tau} = \frac{2\pi}{\hbar}\sum_{\vec{q}} g\frac{e^{-2qD}}{q}
| u^{\dagger}_{\vec{l}-\vec{q}}u_{\vec{l}} |^2
\delta (E^{el}_{\vec{l}-\vec{q}} +\hbar \omega_s - E^{el}_{\vec{l}})
\label{eq:ScatterRateBL}
\end{aligned}
\end{equation}
Eq.~\ref{eq:ScatterRateBL} is the same as Fermi's Golden rule in the context of the materials of our interest. The spin degeneracy factor $g_s$ needs to be multiplied to Eq.~\ref{eq:ScatterRateBL} for completeness. We will insert it in the final expressions of decay rates for the individual materials. The rate calculated in Eq.~\ref{eq:ScatterRateBL} has been normalized by the no. of phonons present in the system. We will use it for all of our subsequent calculations involving the decay rates of the three systems under study. 

\section{The decay rate for BL graphene}
The decay rate for BL graphene is obtained from Eq.~\ref{eq:ScatterRateBL} by replacing $E^{el}_{\vec{l}}$ and $E^{el}_{\vec{l}-\vec{q}}$ by appropriate energy expressions for BL graphene. Also, having replaced $| u^{\dagger}_{\vec{l}-\vec{q}}u_{\vec{l}} |^2$ in Eq.~\ref{eq:ScatterRateBL} by  Eq.~\ref{eq:OverlapBLGFinal}, and $g$ in Eq.~\ref{eq:ScatterRateBL} by Eq.~\ref{eq:ExpressionG}, one obtains the following decay rate for BL graphene in terms of DEMs $q^\prime$ and $l^\prime$.

\begin{equation}             
\begin{aligned}
&\frac{1}{\tau} 
= \frac{\tilde{\epsilon}e^2(m\omega_s)^{\frac{1}{2}}}{\hbar^{\frac{3}{2}}}\int d^2 q^\prime \frac{e^{-2q^\prime \tilde{d}}}{q^ \prime}
\Bigg[1-\frac{{q^\prime}^2\sin^2(\phi^\prime)}{|\vec{l^\prime}-\vec{q^\prime}|^2}\Bigg] \\
& \hspace{40 mm}\delta \Bigg(\frac{{|\vec{l^\prime}-\vec{q^\prime}|}^2}{2} +1 - \frac{{|\vec{l^\prime}|}^2}{2}\Bigg)
\label{eq:ScatterRateBLDimless}
\end{aligned}
\end{equation}
\\
\\
In Eq.~\ref{eq:ScatterRateBLDimless}, the integration w.r.t the $q^\prime$ variable is carried out utilizing properties of $\delta$ function. The details are given in Appendix \ref{app:app2}. Finally, with the spin degeneracy factor $g_s$ inserted, BL graphene decay rate in Eq.~\ref{eq:ScatterRateBLDimless} assumes the following form.  
\begin{equation} 
\begin{aligned}
&\frac{1}{\tau} = \\
&\frac{g_s\tilde{\epsilon}e^2(m\omega_s)^{\frac{1}{2}}}{\hbar^{\frac{3}{2}}} \int_0^{\cos^{-1}\frac{\sqrt{2}}{l^\prime}} d\phi^\prime  
\frac{4e^{-2l^\prime \tilde{d} \cos \phi^\prime}}{L_1} \\
&\Bigg [\cosh (2L_1 \tilde{d}) 
-\frac{\sin^2 \phi^\prime}{{l^\prime}^2-2}
\Bigg( (2{l^\prime}^2 \cos^2 \phi^\prime -2)\cosh (2L_1\tilde{d}) \\
&\hspace{30mm}-2 L_1 l^\prime \sinh(2L_1\tilde{d}) \cos \phi^\prime \Bigg) \Bigg],
\label{eq:BL DeltaPhiIntNotAppend}
\end{aligned}
\end{equation}

where
\begin{equation} 
\begin{aligned}
L_1 \equiv \sqrt{{l^\prime}^2 \cos^2 \phi^\prime-2}
\end{aligned}
\end{equation}

Following is the plot for the numerically evaluated decay rate obtained from Eq.~\ref{eq:BL DeltaPhiIntNotAppend}, $\Bigg ($in units of $\frac{g_s\tilde{\epsilon}e^2(m\omega_s)^{\frac{1}{2}}}{\hbar^{\frac{3}{2}}}$ $\Bigg)$, in the appropriate range of the DEM $l^\prime$, viz., $l^\prime > \sqrt{2}$. 

\begin{figure}[ht]
\begin{center}
\includegraphics[draft=false, bb=0 0 550 550, clip, width=\columnwidth]{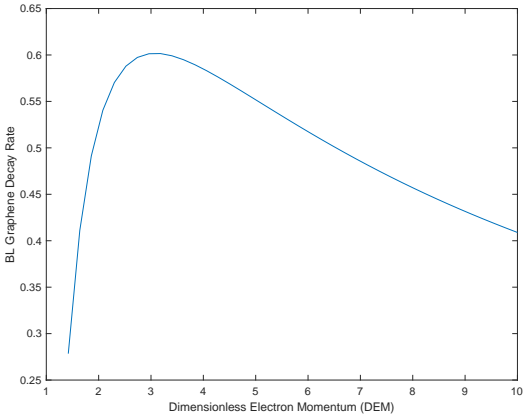}
\caption{Decay Rate $\Bigg ($in units of $\frac{g_s\tilde{\epsilon}e^2(m\omega_s)^{\frac{1}{2}}}{\hbar^{\frac{3}{2}}}$ $\Bigg)$ Versus Dimensionless Electron Momentum (DEM) $l^\prime$ for BL graphene in the allowed range: $l^\prime >\sqrt{2}$.}
\label{fig:BLScatteringVsDimlessL}
\end{center}
\end{figure}
It is seen from Fig.~\ref{fig:BLScatteringVsDimlessL} that the decay rate for BL graphene increases, peaks at a certain value of DEM $l^\prime$ and then falls off as $l^\prime$ further increases. This decay rate, when compared with that of conventional two dimensional polar crystals, agrees in the large DEM range. But the decay rate in conventional two dimensional polar crystals decreases monotonically to zero with the increase of DEM.(This can be shown with the help of a calculation similar to the one carried out by Feynman\cite{Feynman}.) In case of BL graphene, however, the decay rate peaks before falling off.  

\section{Numerical evaluation of polaron energy and the decay rate for SL graphene}

\begin{figure}[ht]
\begin{center}
\includegraphics[draft=false,bb=0 0 550 550, clip, width=\columnwidth]{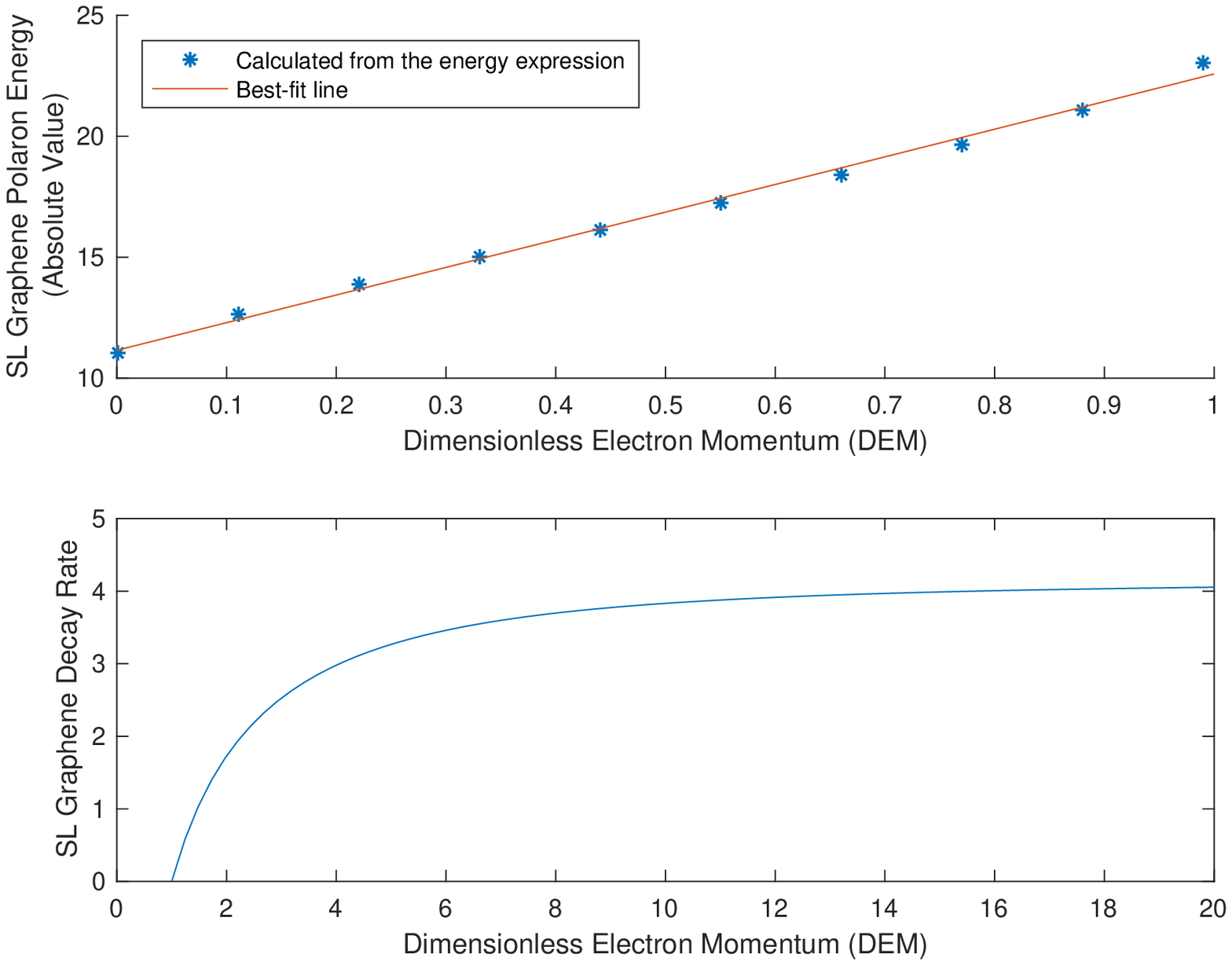}
\caption{Above: Absolute Value of Polaron Energy (in units of $ \frac{g_s\tilde{\epsilon}e^2\omega_s}{4\pi v_F}$) Versus Dimensionless Electron Momentum (DEM) $l^\prime$ for SL graphene in the allowed range: $l^\prime < 1$ \\
Below: Decay Rate (in units of $\frac{g_s\tilde{\epsilon}e^2\omega_s}{\hbar v_F}$) Versus Dimensionless Electron Momentum (DEM) $l^\prime$ for SL graphene in the allowed range: $l^\prime >1$.}
\label{fig:SLEnrgScattering}
\end{center}
\end{figure}

In this section, for SL graphene on a polar substrate, we will compute the polaron energy and the decay rate. SL graphene Hamiltonian is given by $H^0_{SLG} \equiv \hbar v_F\begin{pmatrix}
0 & p_x-ip_y \\
p_x+ip_y & 0,
\end{pmatrix}$.
$v_F$ is the Fermi velocity, and $\vec{p} \equiv (p_x, p_y)$ is the electron wave-vector. The conduction band electron energy of $H^0_{SLG}$ is given by $E = \hbar v_F|\vec{p}| \equiv \hbar v_F p $ and the corresponding electron eigen-spinor, by
\begin{equation}
\begin{aligned}
u_{\vec{p}} =\frac{1}{\sqrt{2}} \begin{pmatrix}
1 \\
e^{i\theta_{\vec{p}}}, 
\end{pmatrix}
\label{eq:GrapheneSpinor}
\end{aligned}
\end{equation}
where $\theta_{\vec{p}} = \atan \frac{p_y}{p_x}$.
\\
\\
In order to obtain polaron energy for SL graphene, the electron eigen-spinors $u_{\vec{l}}$ and $u_{\vec{l}-\vec{q}}$ in Eq.~\ref{eq:EnergyCorrectionFinal} are substituted by appropriate SL graphene eigen-spinors. This can be accomplished by replacing $\vec{p}$ in Eq.~\ref{eq:GrapheneSpinor} by $\vec{l}$ and $\vec{l}-\vec{q}$ respectively. Also, in Eq.~\ref{eq:EnergyCorrectionFinal}, the `non-interacting' electron energies $E^{el}_{\vec{l}}$ and $E^{el}_{\vec{l}-\vec{q}}$ are given by the SL graphene electron energies $\hbar v_F|\vec{l}|$ and $\hbar v_F|\vec{l}-\vec{q}|$ respectively. Hence the polaron energy, as given by Eq.~\ref{eq:EnergyCorrectionFinal} and defined as $\Delta E_{SLG}$ for SL graphene, assumes the following form.

\begin{equation}
\begin{aligned}
& \Delta E_{SLG} = -\sum_{\vec{q}} g\frac{e^{-2qD}}{q}
\frac{| u^{\dagger}_{\vec{l}-\vec{q}}u_{\vec{l}} |^2}
{ \hbar v_F |\vec{l}- \vec{q}| +\hbar\omega_s -  \hbar v_Fl} 
\label{eq:EnergyGrapheneSpinorFinal}
\end{aligned}
\end{equation}
$|{u^{\dagger}}_{\vec{l}-\vec{q}}u_{\vec{l}} |^2$ in Eq.~\ref{eq:EnergyGrapheneSpinorFinal} for SL graphene can be shown to be equal to $\frac{1}{2}\Bigg(1+\frac{1}{|\vec{l}-\vec{q}|}{\Bigg [l-\frac{\vec{l}\cdot \vec{q}}{|\vec{l}|}\Bigg ]}\Bigg)$ with the help of the SL graphene eigen-spinors. Hence one finally obtains from Eq.~\ref{eq:EnergyGrapheneSpinorFinal}

\begin{equation}
\begin{aligned}
& \Delta E_{SLG} = -\sum_{\vec{q}} g\frac{e^{-2q D}}{q}
\frac{\frac{1}{2}\Bigg(1+\frac{1}{|\vec{l}-\vec{q}|}{\Bigg [l-\frac{\vec{l}\cdot \vec{q}}{|\vec{l}|}\Bigg ]}\Bigg)}
{\hbar v_F |\vec{l}-\vec{q}| +\hbar \omega_s -  \hbar v_F l} 
\label{eq:EnergyGrapheneFinalCoordinateIndependent}
\end{aligned}
\end{equation}
As we did in case of BL graphene, we replace $\sum_{\vec{q}}$ by $A\int \frac{d^2\vec{q}}{(2\pi)^2}\equiv A\int_0^\infty qdq \int_0^{2\pi} \frac{d\phi^{\prime}}{(2\pi)^2}$, where $\phi^{\prime}$ is the angle between $\vec{l}$ and $\vec{q}$. We also introduce the dimensionless momenta (DEMs) $\vec{l^\prime} = \frac{v_F}{\omega_S}\vec{l}$, and $\vec{q^\prime} = \frac{v_F}{\omega_S}\vec{q}$. Finally replacing $g$ by Eq.~\ref{eq:ExpressionG} and inserting the spin degeneracy factor $g_s$, Eq.~\ref{eq:EnergyGrapheneFinalCoordinateIndependent} assumes the following form in terms of the dimensionless variables $l^\prime$ and $q^\prime$, and the angle $\phi^{\prime}$ between them. 

\begin{equation}
\begin{aligned}
\Delta E_{SLG} =&  -\frac{g_s\tilde{\epsilon}e^2\omega_s}{4\pi v_F}\int_0^\infty dq^\prime e^{-2q^\prime \tilde{d}} \int_0^{2\pi} d\phi^{\prime} \\
&\frac{1+\frac{1}{\sqrt{{l^\prime}^2 + {q^\prime}^2 - 2l^\prime q^\prime \cos \phi^{\prime}}}{[l^\prime-q^\prime\cos \phi^{\prime}]}}
{\sqrt{{l^\prime}^2 + {q^\prime}^2 - 2l^\prime q^\prime \cos \phi^{\prime}} +1 -  l^\prime} 
\label{eq:EnergyGrapheneFinalDimless}
\end{aligned}
\end{equation},
where $\tilde{d} \equiv \frac{\omega_s D}{v_F}$, a dimensionless constant. As was done in case of BL graphene, we consider the SL graphene on $SiC$ substrate\cite{PolGAndBLG}. Using $\omega_s = 116$ meV (The surface phonon frequency of the substrate $SIC$), $D=4\AA$ and $v_F = 10^6$ meters/sec, we obtain $\tilde{d}=.07$. We will recourse to numerical methods for evaluating Eq.~\ref{eq:EnergyGrapheneFinalDimless}. It is noted that as long as $l^\prime <1$, the denominator in  Eq.~\ref{eq:EnergyGrapheneFinalDimless} is always positive for all values of $q^\prime$ and $\phi^{\prime}$, thereby guaranteeing $\Delta E_{SLG}$ will not be singular. $l^\prime <1$ sets the upper limit of DEM $l^\prime$ for which polaron formation is possible for SL graphene. One detects an apparent singularity in the expression $\frac{1}{\sqrt{{l^\prime}^2 + {q^\prime}^2 - 2l^\prime q^\prime \cos \phi^{\prime}}}$, in the numerator of the integrand of Eq.~\ref{eq:EnergyGrapheneFinalDimless}. This expression becomes infinite at $q^\prime = l^\prime$, and $\phi^{\prime} = 0$. But since the factor $[l^\prime-q^\prime\cos \phi^{\prime}]$, which multiplies the above mentioned expression, goes to zero at $q^\prime = l^\prime, \phi^{\prime} = 0$, it's not a true singularity. For $l^\prime > 1$, the denominator in Eq.~\ref{eq:EnergyGrapheneFinalDimless} can actually become $0$, rendering $\Delta E_{SLG}$ singular, for some $q^\prime$ and $\phi^{\prime}$. Hence $l^\prime > 1$ will correspond to decay, the rate of which will be given by the general expression Eq.~\ref{eq:ScatterRateBL}. From Eq.~\ref{eq:ScatterRateBL}, the decay rate for SL graphene is calculated by replacing $g$ by Eq.~\ref{eq:ExpressionG}, and $E^{el}_{\vec{l}-\vec{q}}$, $E^{el}_{\vec{l}}$ and  $| u^{\dagger}_{\vec{l}-\vec{q}}u_{\vec{l}} |^2$ by suitable expressions for SL graphene. The last three quantities mentioned in the last line were evaluated while obtaining an expression for $\Delta E_{SLG}$. The decay rate, thus evaluated and expressed in terms of DEMs $q^\prime$ and $l^\prime$, assumes the following form.

\begin{equation}
\begin{aligned}    
\frac{1}{\tau} = & \frac{\tilde{\epsilon}e^2\omega_s}{2 \hbar v_F}\int_0^\infty dq^\prime \int d\phi^{\prime} e^{-2q^\prime \tilde{d}} \\
&\Bigg [1+\frac{l^\prime-q^\prime \cos \phi^{\prime}}{\sqrt{{l^\prime}^2 + {q^\prime}^2 - 2l^\prime q^\prime \cos \phi^{\prime}}}\Bigg] \\
&\delta \Bigg( \sqrt{{l^\prime} ^2 + {q^\prime}^2 - 2l^\prime q^\prime \cos \phi^{\prime}} +1 -  l^\prime \Bigg)
\label{eq:EnergyLossSLG}
\end{aligned}
\end{equation} 
The integration w.r.t $q^\prime$ in Eq.~\ref{eq:EnergyLossSLG} can be carried out utilizing the properties of $\delta$ function. The details of the calculation is given in Appendix \ref{app:app3}. Finally, inserting the spin degeneracy factor $g_s$, the following expression for SL graphene decay rate is obtained from Eq.~\ref{eq:EnergyLossSLG}.
\begin{equation} 
\begin{aligned}
\frac{1}{\tau} = &\frac{g_s\tilde{\epsilon}e^2\omega_s}{\hbar v_F} \int_0^{\cos^{-1}\sqrt{\frac{2l^\prime-1}{{l^\prime}^2}}} d\phi^{\prime} \\  
&\frac{2e^{-2l^{\prime} \tilde{d} \cos \phi^{\prime}}}{L_2} 
\Bigg [(2l^\prime -1) \cosh (2 L_2 \tilde{d}) \\
&-\cos \phi^{\prime} \Bigg(l^\prime \cos \phi^{\prime} \cosh (2 L_2 \tilde{d})
-L_2 \sinh (2 L_2 \tilde{d}) \Bigg) \Bigg],
\label{eq:SL DeltaPhiIntNotAppend}
\end{aligned}
\end{equation}

where
\begin{equation} 
\begin{aligned}
L_2 \equiv \sqrt{{l^\prime}^2 \cos^2 \phi^{\prime}-(2l^\prime -1)}
\end{aligned}
\end{equation}
Fig.~\ref{fig:SLEnrgScattering} shows the plots of the absolute value of polaron energy and decay rate of SL graphene w.r.t to DEM $l^\prime$. The polaron energy and the decay rate are evaluated using Eqs. ~\ref{eq:EnergyGrapheneFinalDimless} and ~\ref{eq:SL DeltaPhiIntNotAppend} respectively.
From Fig.~\ref{fig:SLEnrgScattering} it is seen that the polaron energy for SL graphene changes more or less linearly with dimensionless electron momentum (DEM) for the allowed range of the DEM, i.e., $l^\prime < 1$. For $l^\prime > 1$, the decay rate initially goes up with the increase in DEM, and then flattens. This is quite different from the decay rate pattern of BL graphene, as given by Fig.~\ref{fig:BLScatteringVsDimlessL}. Unlike SL graphene, the decay rate for BL graphene falls off for large values of DEM.
\\
\\
Whereas the decay rate for SL graphene is in stark contrast with the decay rate of BL graphene, the polaron energy of SL graphene shares a striking similarity with the polaron energy of BL graphene for small $l^\prime$. Despite the fact that that SL and BL graphene have very different electronic energy momentum dispersion relationship in the absence of electron-phonon interaction, the two systems behave rather similarly so far as the polaron-energy in the small momentum region is concerned. Incidentally it can be mentioned that the decay rate for the SL graphene polarons is quite similar to the decay rate for polarons in conventional three dimensional polar crystals with quadratically dispersing electrons\cite{Feynman}.

\section{Polaron Energy and Decay Rate for Semi-Dirac on a Polar Substrate}
The energy momentum dispersion relation for a semi-Dirac electron is given by $E =\pm\sqrt{\frac{\hbar^4 p{_x}^4}{4m^2} + \hbar^2 {v_F}^2 p{_y}^2}$, the positive and the negative signs corresponding to the conduction and the valence bands respectively\cite{BanerjeeSd1, BanerjeeSd2, Montambaux}. $m$ is the mass-parameter and $v_F$ is the velocity parameter. $p_x$ and $p_y$ are the electron wave-vectors along two special directions in the Brillouin zone, viz., $x$ and $y$. Along the $x$ direction, semi-Dirac energy disperses quadratically like an ordinary electron. Hence $x$ is called the non-relativistic direction. Along the $y$ direction, semi-Dirac energy disperses linearly, like graphene. Hence $y$ is called the relativistic direction.
\\
\\
The above-mentioned energy-momentum relationship for semi-Dirac can be derived from more than one Hamiltonian related to each other by unitary transformations. To get the essential physics, keeping the computations as simple as possible, we will use the following form of the `non-interacting' semi-Dirac Hamiltonian as given by $H^{SD}_{0} \equiv \begin{pmatrix}
0 & \frac{\hbar^2 p_x^2}{2m}-i\hbar v_F p_y \\
\frac{\hbar^2 p_x^2}{2m}+i\hbar v_Fp_y &  0
\end{pmatrix}$.
The electron eigenstate of $H^{SD}_0$ is given by  
\begin{equation}
\begin{aligned}
u_{\vec{p}} =\frac{1}{\sqrt{2}} \begin{pmatrix}
1 \\
e^{i\theta_{\vec{p}}} 
\end{pmatrix}
\label{eq:SdSpinor}
\end{aligned}
\end{equation}
where, 
\begin{equation}
\theta_{\vec{p}} = \atan(\frac{mv_F}{\hbar}\frac{2p_y}{p_x^2})
\label{eq:thetaSD}
\end{equation}
Next, to compute the polaron energy for semi-Dirac, the spinors $u_{\vec{l}}$ and $u_{\vec{l}-\vec{q}}$ in Eq.~\ref{eq:EnergyCorrectionFinal} are replaced by the semi-Dirac spinors, as given by Eq.~\ref{eq:SdSpinor}, using $\vec{l}$ and $\vec{l}-\vec{q}$ for $\vec{p}$ respectively. The `non-interacting' energies $E^{el}_{\vec{l}}$ and $E^{el}_{\vec{l}-\vec{q}}$ in Eq.~\ref{eq:EnergyCorrectionFinal} for semi-Dirac are given by $\sqrt{\frac{\hbar^4{l_x}^4}{4} + \hbar^2v_F^2{l_y}^2}$ and $\sqrt{\frac{\hbar^4(l_x-q_x)^4}{4m^4}+ \hbar^2v_F^2(l_y-q_y)^2}$ respectively. Using the above-mentioned quantities, the polaron energy, as given by Eq.~\ref{eq:EnergyCorrectionFinal} and defined as $\Delta E^{SD}$ for semi-Dirac, assumes the following form.

\begin{equation}
\begin{aligned}
& \Delta E^{SD} = -\sum_{\vec{q}} \frac{ge^{-2qD}}{2q} \\
& \frac{1+\cos(\atan(\frac{mv_F}{\hbar}\frac{2l_y}{l_x^2})-\atan\Bigg(\frac{mv_F}{\hbar}\frac{2(l_y - q_y)}{(l_x - q_x)^2}\Bigg))}
{{{\sqrt{\frac{\hbar^4(l_x-q_x)^4}{4m^4}+ \hbar^2v_F^2(l_y-q_y)^2} +\hbar\omega_s} - \atop \hspace{45mm} {\sqrt{\frac{\hbar^4l{_x}^4}{4} + \hbar^2v_F^2l{_y}^2}}}} 
 \label{eq:EnergySDxy}
\end{aligned}
\end{equation}
We will convert the wave-vectors $\vec{l}$ and $\vec{q}$ appearing in Eq.~\ref{eq:EnergySDxy} into dimensionless momenta variables (DEMs) in a spirit similar to the one adopted for SL and BL graphene. But this time, due to the anisotropic nature of the semi-Dirac dispersion, we will scale the $x$ and the $y$ components of the wave-vectors differently, i.e., divide the $x$ and the $y$ components of the wave-vectors by different constants to render them dimensionless. For the wave-vector $\vec{q}(\equiv (q_x, q_y))$ we define the DEMs $q_x^\prime = \sqrt{\frac{\hbar}{m\omega_S}} q_x$, and $q_y^\prime = \frac{v_F}{\omega_S} q_y$, and an exactly similar set of DEMs for the wave-vector $\vec{l}\equiv (l_x, l_y)$. The standard replacements of $g$ by Eq.~\ref{eq:ExpressionG}, and $\sum_{\vec{q}}$ by $A\int^{\infty}_{-\infty} \int^{\infty}_{-\infty}\frac{dq_x dq_y}{(2\pi)^2}$ are done in Eq.~\ref{eq:EnergySDxy}. Finally, with the insertion of the spin degeneracy factor $g_s$, Eq.~\ref{eq:EnergySDxy} assumes the following form in terms the above-mentioned DEMs. 

\begin{equation}
\begin{aligned}
& \Delta E^{SD} \\
&= -\frac{g_s\tilde{\epsilon}e^2}{4\pi}\sqrt{\frac{m\omega_S}{\hbar}} \int dq_x^\prime dq_y^\prime 
\frac{e^{-2\tilde d_{SD}\sqrt{{q_y^\prime}^2+\kappa {q_x^\prime}^2}}}{\sqrt{{q_y^\prime}^2+\kappa {q_x^\prime}^2}}\frac{W}{R}, 
 \label{eq:EnergySDxyDimless}
\end{aligned}
\end{equation}
where 
where $W$ and $R$ are given by 
\begin{subequations}
\begin{equation}
\begin{aligned}
W = 1+\cos(\atan \frac{2l_y^\prime}{{l_x^\prime}^2}-\atan \frac{2(l_y^\prime - q_y^\prime)}{(l_x^\prime - q_x^\prime)^2})
\label{eq:SDEnergyNum}
\end{aligned}
\end{equation}

\begin{equation}
\begin{aligned}
R = \sqrt{\frac{(l_x^\prime-q_x^\prime)^4}{4}+ (l_y^\prime-q_y^\prime)^2} +1 - \sqrt{\frac{{l_x^\prime}^4}{4} + {l_y^\prime}^2}
\label{eq:SDEnergyDenom}
\end{aligned}
\end{equation}
\label{eq:SDEnergyNumDenom}
\end{subequations}
\\
\\
In Eq.~\ref{eq:EnergySDxyDimless}, $\tilde d_{SD}$ is a dimensionless quantity given by $\tilde d_{SD} = \frac{D\omega_S}{v_F}$. $\kappa$ is also a dimensionless constant given by $\kappa \equiv \frac{mv_F^2}{\hbar \omega_S}$. For $SIC$ substrate the surface phonon frequency $\omega_S=116$ meV. Replacing $m$ by the electron mass, $v_F$ by $10^6$m/sec, and $D$ by $15 \AA$, one obtains $\tilde d_{SD} \approx .26$ and $\kappa \approx 49$. 
\\
\\
Next, we will study divergences in the integrand of Eq.~\ref{eq:EnergySDxyDimless}, since that will set an upper limit for DEM for which polaron formation is possible. First of all, it appears from Eq.~\ref{eq:EnergySDxyDimless} that the integrand on the right side diverges for $q_x^\prime = 0, q_y^\prime = 0$ due to the presence of the factor $\frac{e^{-2\tilde d_{SD}\sqrt{{q_y^\prime}^2+\kappa {q_x^\prime}^2}}}{\sqrt{{q_y^\prime}^2+\kappa {q_x^\prime}^2}}$. But by studying the integrand in the neighborhood of the point $(q_x^\prime = 0, q_y^\prime = 0)$, one can convince oneself that it stays integrable, nonetheless, as long as $\kappa$ is greater than $1$.
\\
\\
The real divergence in Eq.~\ref{eq:EnergySDxyDimless} comes from $R$. As long as $R$ is not $0$, there is no divergence in the expression of $\Delta E^{SD}$ . From the expression of $R$ given by Eq.~\ref{eq:SDEnergyDenom}, it can be seen that $R$ is guaranteed to be greater than $0$, if $\sqrt{\frac{{l_x^\prime}^4}{4} + {l_y^\prime}^2}$ is less than $1$. Defining $\gamma$ as the angle that DEM $\vec{l^\prime} \equiv (l^\prime_x, l^\prime_y)$ makes w.r.t the $x$-axis, and writing $l_x^\prime = l^\prime \cos \gamma$ and $l_y^\prime = l^\prime \sin \gamma$, $\sqrt{\frac{{l_x^\prime}^4}{4} + {l_y^\prime}^2}$ can be expressed as $\sqrt{\frac{{l^\prime}^4\cos^{4}\gamma}{4} + {l^\prime}^2\sin^{2}\gamma}$. It can be shown with a little algebra that the above-mentioned quantity is less than 1, if  $l^\prime$ satisfies the following criterion. 

\begin{equation}
\begin{aligned}
l^\prime < \sqrt{\frac{2}{\cos^{4}\gamma}(-\sin^{2}\gamma+\sqrt{\sin^{4}\gamma + \cos^{4}\gamma})}
\label{eq:SDMomentumRange}
\end{aligned}
\end{equation}

\begin{figure}[ht]
\begin{center}
\includegraphics[draft=false,bb=0 0 550 550, clip, width=\columnwidth]{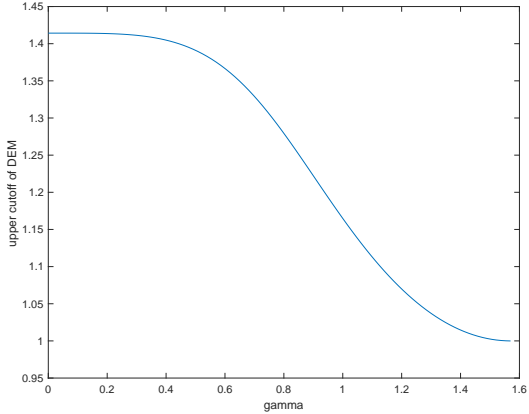}
\caption{Upper Cutoff of the Dimensionless Electron Momentum (DEM) $l^\prime$ versus angle $\gamma$ for Semi-Dirac.}
\label{fig:SDMomntmVsAng}
\end{center}
\end{figure}

\begin{figure}[ht]
\begin{center}
\includegraphics[draft=false,bb=0 0 550 550, clip, width=\columnwidth]{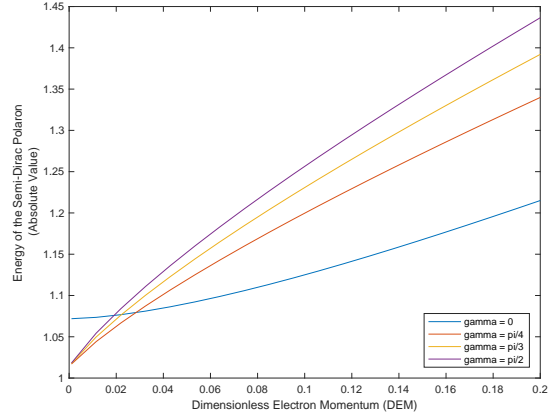}
\caption{Absolute Value of the Energy $\Bigg($in the units of $\frac{g_s\tilde{\epsilon}e^2}{4\pi}\sqrt{\frac{m\omega_S}{\hbar}}\Bigg)$ Vs Dimensionless Electron Momentum (DEM) $l^\prime$ for various $\gamma$s for Semi-Dirac.}
\label{fig:SDEnrgyVsMomntmAngs}
\end{center}
\end{figure}

\begin{figure}[ht]
\begin{center}
\includegraphics[draft=false,bb=0 0 550 550, clip, width=\columnwidth]{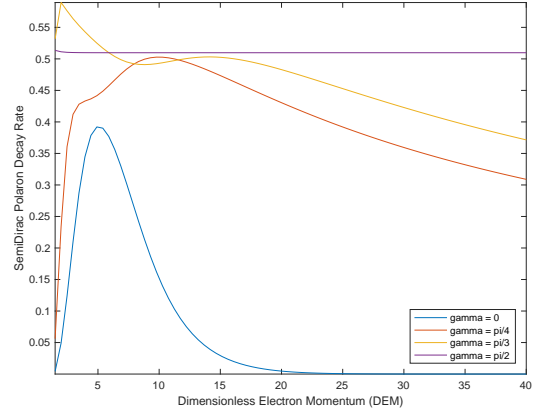}
\caption{Decay Rate $\Bigg($ in the units of $\frac{g_s\tilde{\epsilon}e^2(m\omega_s)^\frac{1}{2}}{2\hbar^{\frac{3}{2}}}\Bigg)$ Vs Dimensionless Electron Momentum (DEM) $l^\prime$ for various $\gamma$s for Semi-Dirac.}
\label{fig:SDScattering}
\end{center}
\end{figure}

Inequality \ref{eq:SDMomentumRange} sets an upper-bound for DEM $l^\prime$, for which $\Delta E^{SD}$ is well defined and hence polaron formation is possible. It is seen from inequality \ref{eq:SDMomentumRange} that this upper limit for the semi-Dirac system is a function of the angle $\gamma$ that $\vec{l^\prime}$ makes with $x$-axis. This is due to the anisotropic nature of the energy-momentum relation of a semi-Dirac system. The upper-bound of $l^\prime$, i.e., the right side of the inequality ~\ref{eq:SDMomentumRange}, can be proven to vary from $\sqrt{2}$ to $1$. The upper-bound of $l^\prime$ assumes the value $\sqrt{2}$ when $\gamma = 0$, corresponding to the electron momentum being in the $x$ or the `non-relativistic' direction. The upper-bound of $l^\prime$ is $1$ when $\gamma = \frac{\pi}{2}$, corresponding to the electron momentum being in the $y$ or the `relativistic' direction. For any intermediate angle, the upper-bound is in between these two limits. There is no need to consider $\gamma$ beyond $\frac{\pi}{2}$, since in inequality ~\ref{eq:SDMomentumRange} only the even powers of quantities like $\sin\gamma$, and  $\cos\gamma$ appear. In Fig.~\ref{fig:SDMomntmVsAng}, the angular dependence of the upper bound of the DEM $l^\prime$ for which polaron formation is possible in a semi-Dirac system, is plotted against $\gamma$.    
It can be seen from the figure that the upper-bound of the DEM varies monotonically with angle $\gamma$ between the two extreme limits $1$ and $\sqrt{2}$, as mentioned before. 
\\
\\
Next, the absolute value of the polaron energy as given by Eq.~\ref{eq:EnergySDxyDimless} for the semi-Dirac system, is plotted, in the units of $\frac{g_s\tilde{\epsilon}e^2}{4\pi}\sqrt{\frac{m\omega_S}{\hbar}}$ in Fig.~\ref{fig:SDEnrgyVsMomntmAngs} for small DEMs.  
It is seen that so far as the semi-Dirac polaron energy goes, there is a stark difference between the `non-relativistic' direction corresponding to $\gamma = 0$, and any other direction. There is a gap in the energy values between the non-relativistic direction and other directions when the DEM $l^\prime = 0$. For all the directions excepting the `non-relativistic' one, the polaron energies tend to the same unique value when the DEM $l^\prime$ approaches the value $0$. This exotic limiting behavior of polaron energy puts the semi-Dirac system in a very different category from other materials including SL and BL graphene.  
\\
\\
Next, we discuss whether the semi-Dirac polaron energy disperses linearly with DEM. This is not as straightforward as it was for the cases of SL and BL graphene, because, unlike them, DEM in semi-Dirac is scaled differently for $x$ and $y$ directions. For an arbitrary angle $\gamma$, even if, say, the semi-Dirac polaron energy looked linear in DEM $l^\prime$ from the plot, we could not have concluded that the energy is linear in the electron-momentum. DEM and the actual electron momentum for the semi-Dirac system aren't quite equivalent, due to the unequal scaling of the semi-Dirac electron-momentum in the two different directions. 
\\
\\
The exception to this happens when the electron momentum has either only the $x$-component or only the $y$-component, i.e., the electron moves either in the `non-relativistic' or in the `relativistic' direction. Then, because there is just one type of scaling, there exists a perfect equivalence between DEM and the electron momentum once again. From the plot of Fig.~\ref{fig:SDEnrgyVsMomntmAngs}, the polaron energies in these two special directions, corresponding to $\gamma = 0$, and $\gamma = \frac{\pi}{2}$, appear to not be linear for small momentum. This behavior is a departure from the linear nature of the polaron energy-momentum dispersion in both the SL and the BL graphene for small momentum. This shows that semi-Dirac, although resembling SL and the BL graphene along two special directions from the point of view of the non-interacting electron-energy, behaves very differently from those materials so far as its polaron energies in those directions are concerned. 
\\
\\
Finally, we consider the decay rate when the semi-Dirac electron momentum does not satisfy the inequality given by ~\ref{eq:SDMomentumRange}. In the decay rate formula given by Eq.~\ref{eq:ScatterRateBL}, we will replace $g$ by Eq.~\ref{eq:ExpressionG} as usual; and $E^{el}_{\vec{l}-\vec{q}}$ and $E^{el}_{\vec{l}}$ by the `non-interacting' energy expressions for semi-Dirac electrons corresponding to the wave-vectors $\vec{l}-\vec{q}$ and $\vec{l}$ respectively. Also, in Eq.~\ref{eq:ScatterRateBL} we replace $| u^{\dagger}_{\vec{l}-\vec{q}}u_{\vec{l}} |^2$ by $W$, as given by Eq.~\ref{eq:SDEnergyNum}, and the argument of the $\delta$ function by $R$, as given by Eq.~\ref{eq:SDEnergyDenom}. Finally, employing the definitions of DEMs $\vec{l^\prime}$ and $\vec{q^\prime}$, the decay rate, as per Eq.~\ref{eq:ScatterRateBL} takes the following form for the semi-Dirac polaron. 
\begin{equation}
\begin{aligned}    
\frac{1}{\tau} =&\frac{\tilde{\epsilon}e^2(m\omega_s)^\frac{1}{2}}{2\hbar^{\frac{3}{2}}}\int dq^\prime_xdq^\prime_y \\
 &\frac{e^{-2\tilde d_{SD}\sqrt{{q_y^\prime}^2+\kappa {q_x^\prime}^2}}}{\sqrt{{q_y^\prime}^2+\kappa {q_x^\prime}^2}}W\delta(R),
\label{eq:ScatterEnergySD}
\end{aligned}
\end{equation} 
where $W$ and $R$ are given by Eq.~\ref{eq:SDEnergyNumDenom}. Utilizing the properties of $\delta$ function, the integration w.r.t the $q^\prime_y$  variable in Eq.~\ref{eq:ScatterEnergySD} can be carried out. The details are given in Appendix \ref{app:app4}. Finally inserting the spin degeneracy factor $g_s$, we obtain from Eq.~\ref{eq:ScatterEnergySD},
\begin{equation}
\begin{aligned}
&\frac{1}{\tau} 
=\frac{g_s\tilde{\epsilon}e^2(m\omega_s)^\frac{1}{2}}{2\hbar^{\frac{3}{2}}}\\
&\bigintss_{l^\prime_x - \sqrt{2}\sqrt{ \sqrt{\frac{1}{4}{l^\prime_x}^4+{l^\prime_y}^2}-1}}^{l^\prime_x + \sqrt{2}\sqrt{\sqrt{\frac{1}{4}{l^\prime_x}^4+{l^\prime_y}^2}-1}} dq^\prime_x
\frac{\sqrt{\frac{1}{4}{l^\prime_x}^4+{l^\prime_y}^2}-1}{L_3} \\
& \Bigg [ \frac{e^{-2\tilde{d}\sqrt{Q+2l^\prime_y L_3}}}
{\sqrt{Q+2l^\prime_y L_3}}
\Bigg(1+\cos(\atan \frac{2l_y^\prime}{{l_x^\prime}^2}+\atan \frac{2 L_3}{(l_x^\prime - q_x^\prime)^2})\Bigg) \\
& \hspace{40mm} + \\
&\frac{e^{-2\tilde{d}\sqrt{Q-2l^\prime_y L_3}}}
{\sqrt{Q-2l^\prime_y L_3}}
\Bigg(1+\cos(\atan \frac{2l_y^\prime}{{l_x^\prime}^2}-\atan \frac{2 L_3}{(l_x^\prime - q_x^\prime)^2})\Bigg)  \Bigg],
\label{eq:SDDeltaFreeScatterNotAppend}
\end{aligned}
\end{equation}
 \\
where $L_3 \equiv \sqrt{ \Bigg(\sqrt{\frac{1}{4}{l^\prime_x}^4+{l^\prime_y}^2}-1\Bigg)^2-\frac{1}{4}(l^\prime_x-q^\prime_x)^4}$, and $Q \equiv \kappa {q_x^\prime}^2 + {l^\prime_y}^2 + L^2_3$.
\\
\\
In Fig.~\ref{fig:SDScattering} the decay rate, numerically evaluated from Eq.~\ref{eq:SDDeltaFreeScatterNotAppend}, is plotted in the units of $\frac{g_s\tilde{\epsilon}e^2(m\omega_s)^\frac{1}{2}}{2\hbar^{\frac{3}{2}}}$, as a function of DEM for various $\gamma$s. For all the plots the lower limit of the DEM has been chosen to be $\sqrt{2}$, which is outside the `polaron-formation region' for all the values of $\gamma$ as per Fig.~\ref{fig:SDMomntmVsAng}.
From Fig.~\ref{fig:SDScattering} it can be seen that the decay rate changes with DEM differently for different $\gamma$s. In Fig.~\ref{fig:SDScattering} the plot of the decay rate for $\gamma = 0$ (the non-relativistic direction) vs DEM is similar to the plot of the BL graphene decay rate vs DEM, as given by Fig.~\ref{fig:BLScatteringVsDimlessL}. This behavior is commensurate with the fact that the semi-Dirac energy momentum dispersion of non-interacting electrons along $\gamma = 0$ reduces to that of BL graphene.
\\
\\
Next we compare the plot of the decay rate vs DEM for $\gamma = \frac{\pi}{2}$ in Fig.~\ref{fig:SDScattering} with the decay rate vs. DEM plot for SL graphene, given by Fig.~\ref{fig:SLEnrgScattering}. This is of interest since the semi-Dirac energy momentum dispersion of non-interacting electrons along $\gamma = \frac{\pi}{2}$ reduces to that of SL graphene. It is seen that the decay rate stays more or less constant with DEM for semi-Dirac along $\gamma = \frac{\pi}{2}$, whereas for SL graphene the decay rate, after initially increasing with DEM, flattens out. It is claimed that the aforementioned two plots are nonetheless similar. The absence of the initial increase of the decay rate in case of semi-Dirac is attributed to the large value of $\kappa$ used in the numerical evaluation of Eq.~\ref{eq:SDDeltaFreeScatterNotAppend}. It has been checked that with smaller $\kappa$'s one can actually observe the decay rate increasing before flattening out in a similar vein along the SL graphene decay rate. Hence, so far as the decay rate goes, semi-Dirac behaves as BL graphene or SL graphene depending on whether the electron-momentum is aligned along the non-relativistic or the relativistic direction. This is commensurate with the fact that the semi-Dirac energy momentum dispersion of non-interacting electrons reduces to that of BL graphene (SL graphene) along non-relativistic (relativistic) direction. In Fig.~\ref{fig:SDScattering}, as for the values of $\gamma$ which are in between $0$ and $\frac{\pi}{2}$, the decay rate vs DEM plots are similar to that of $\gamma = 0$ in the sense that the decay rate falls off for sufficiently large values of DEM. Hence the relativistic direction ($\gamma =\frac{\pi}{2}$) stands out w.r.t the decay rate of the semi-Dirac system. 

\section{Summary}
In this paper the polaron dynamics for the three two dimensional semi-metals, viz., BL and SL graphene, and semi-Dirac has been studied. The materials are assumed to be grown on polar substrates. Both the polaron energy and the decay rate are calculated for all the three systems. A novel finding of polaron energy dispersing linearly with small electron momenta for BL graphene, has been presented. This result, which has been derived analytically, is very different from the usual small-momentum quadratic energy momentum dispersion relation of polarons in polar crystals.  The polaron energy for SL graphene, evaluated numerically, has been shown to vary approximately linearly. The decay rates vary quite differently for the BL and SL graphene. While for the former the decay rate falls off with large elecron-momenta, for the latter it assumes a constant value. For semi-Dirac it has been observed when the electron momentum goes to zero, the polaron energy assumes two distinctly different values. The values differ depending on whether the electron momentum is approaching zero from the non-relativistic direction or from any other directions. This direction-dependent non-uniqueness of polaron energy for vanishing electron momentum is an unique feature of semi-Dirac, not shared by the other two systems. In other respects semi-Dirac shares features with SL and BL graphene.

\appendix
\section{Derivation of the integrals $J_1, J_2$ and $J_3$ for small dimensionless electron momentum (DEM) $l^\prime$}\label{app:app1}
All the three integrals $J_1$, $J_2$ and $J_3$, as given by Eq.~\ref{eq:JOneTwo}, are evaluated by carrying out the integration w.r.t the $\phi^\prime$ variable first, followed by integration w.r.t the $q^\prime$ variable. The $\phi^\prime$ integration will be performed by going to the complex plane and then using the techniques of complex analysis. A complex variable $z\equiv e^{i\phi^\prime}$, describing a circle of unit radius in the complex plane, is introduced to replace $\cos \phi^\prime$ and $\sin \phi^\prime$ appearing in Eq.~\ref{eq:JOneTwo} by $z+\frac{1}{z}$ and $\frac{1}{2i}(z-\frac{1}{z})$ respectively. Also, $d\phi^\prime$ will be replaced by $\frac{dz}{iz}$. Thus the $\phi^\prime$ integrals will be converted into the integrals w.r.t the complex variable $z$, which will then be evaluated by Cauchy-residue theorem of the complex variables. 

\subsection{Evaluation of $J_1$}
\begin{figure}[ht]
\begin{center}
\includegraphics[draft=false,bb=0 0 550 550, clip, width=\columnwidth]{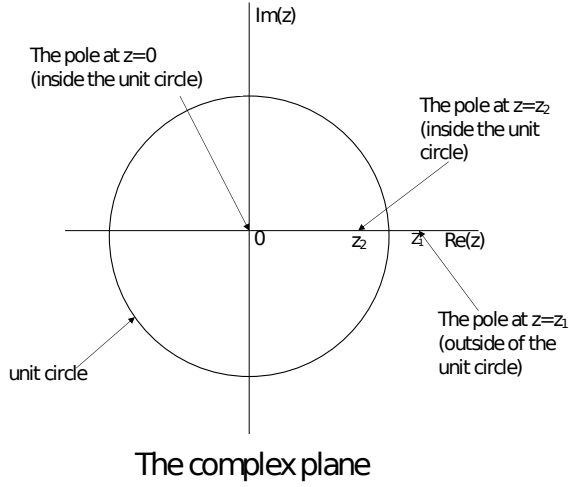}
\caption{The unit circle in the complex plane showing the poles. The poles $z=0$ and $z=z_2$ are inside the unit circle.}
\label{fig:ComplexUnitCir}
\end{center}
\end{figure}

Following the above-mentioned substitutions, the $\phi^\prime$ integral in $J_1$ in Eq.~\ref{eq:JOneQPhi} can be replaced by a complex variable integral resulting in the following expression for $J_1$.  
\begin{equation}
\begin{aligned}
& J_1 = \int_0^\infty dq^\prime \frac{q^\prime e^{-2q^\prime \tilde{d}}}{4l^\prime({q^\prime}^2+2)}\oint dz f(z),
\label{eq:JOneQIntContour}
\end{aligned}
\end{equation}

where 
\begin{equation}
\begin{aligned}
& f(z) = -i\frac{(z^2-1)^2}{z^2(z^2-\frac{{l^\prime}^2+{q^\prime}^2}{l^\prime q^\prime}z+1)},
\label{eq:JOneContour}
\end{aligned}
\end{equation}
\\
\\
The integral $\oint dz f(z)$ in Eq.~\ref{eq:JOneQIntContour} represents contour integration w.r.t the complex variable $z$, the contour being an unit circle in a complex plane as given in Fig.~\ref{fig:ComplexUnitCir}. The contour integral can be evaluated by the standard residue calculus, i.e., finding the residues at the singularities of the function $f(z)$ inside the contour, adding them up and then multiplying the sum by the factor $2\pi i$. 

\begin{equation}
\oint dz f(z) = 2\pi i \sum_i (\text{Res}(f(z)))\Bigg|_{z = z_i} 
\label{eq:ResThm}
\end{equation}
$\text{Res}$ is the short form for `residue'. $z_i$'s are the poles or the singularities of the function $f(z)$, which are inside the unit circle in the complex plane. The singularities of $f(z)$, as can be seen from Eq.~\ref{eq:JOneContour}, are $z=0$ and the roots of the equation $z^2-\frac{{l^\prime}^2+{q^\prime}^2}{l^\prime q^\prime}z+1 = 0$. The roots are
\begin{equation}
\begin{aligned}
z_1 = \frac{1}{2l^\prime q^\prime}\Bigg[{l^\prime}^2+{q^\prime}^2+\abs{{l^\prime}^2-{q^\prime}^2}\Bigg] \\
z_2 = \frac{1}{2l^\prime q^\prime}\Bigg[{l^\prime}^2+{q^\prime}^2-\abs{{l^\prime}^2-{q^\prime}^2}\Bigg]
\label{eq:ZOneZTwo}
\end{aligned}
\end{equation}
The naming of $z_1$ and $z_2$ is arbitrary.  
\\
\\
$z=0$ pole of $f(z)$ is obviously inside the unit circle, as shown in Fig.~\ref{fig:ComplexUnitCir}; and it is a pole of order 2. Both the roots $z_1$ and $z_2$ are poles of order $1$, but only $z_2$ lies inside the unit circle, as shown in Fig.~\ref{fig:ComplexUnitCir}. It can be checked readily that this is true regardless of whether $l^\prime < q^\prime$  or $l^\prime > q^\prime$. It's not of importance to consider the case $l^\prime = q^\prime$, for, this particular case will not have any effect on the final expression for $J_1$. The reason for that is as follows. $q^\prime = l^\prime$ being just one point on the $q^\prime$-axis is of measure $0$. Hence, it will not contribute to the integration w.r.t. the $q^\prime$ variable immediately following the $\phi^\prime$ integration in the evaluation of $J_1$. 
\\
\\ 
The residues of $f(z)$ are evaluated at $z = 0$, and $z = z_2$, the only poles inside the unit circle, using the following formulae. $z = z_2$ being a simple pole, $Res (f(z))|_{z=z_2} = \lim_{z \to z_2} (z-z_2)f(z)$. Next, $z=0$ being a pole of order $2$, $Res (f(z))|_{z=0} = \lim_{z \to 0}\frac{d}{dz}\Bigg[z^2f(z)\Bigg]$. Evaluating the residues and subsequently adding them together, after some algebra, one obtains a rather simple result: $\sum_i (Res(f(z)))\Bigg|_{z = z_i} = -2i z_2$. [The algebra involves repeated use of the identity $z_1z_2 = 1$, the validity of which can readily be checked from Eq.~\ref{eq:ZOneZTwo}]. Replacing $\sum_i (Res(f(z)))\Bigg|_{z = z_i}$ by $-2i z_2$ in Eq.~\ref{eq:ResThm}, one obtains the following expression for the contour integral $\oint dz f(z)$.     
\begin{equation}
\begin{aligned}
\oint dz f(z)  & = 4\pi z_2 
\label{eq:ContourIntFinal}
\end{aligned}
\end{equation}
Using Eq.~\ref{eq:ContourIntFinal} in Eq.~\ref{eq:JOneQIntContour}, with $z_2$ given by Eq.~\ref{eq:ZOneZTwo}, one obtains,
\begin{equation}
\begin{aligned}
& J_1 = \frac{\pi}{2 {l^\prime}^2}\Bigg[\int_0^{l^\prime}dq^\prime e^{-2q^\prime \tilde{d}}\frac{2{q^\prime}^2}{{q^\prime}^2+2} \\
&\hspace{40mm}+ \int_{l^\prime}^\infty dq^\prime e^{-2q^\prime \tilde{d}} \frac{2{l^\prime}^2}{{q^\prime}^2+2}\Bigg]
\label{eq:JOneQInt}
\end{aligned}
\end{equation}
\\
\\
There are two $q^\prime$ integrals in Eq.~\ref{eq:JOneQInt}, which is a direct result of the fact that $z_2$, as can be seen from Eq.~\ref{eq:ZOneZTwo}, assumes two different expression for the the two regions: $q^\prime < l^\prime$  or $q^\prime > l^\prime$. For $q^\prime < l^\prime$ $z_2 = 2{q^\prime}^2$, and for $q^\prime > l^\prime$, $z_2 = 2{l^\prime}^2$. What happens at $q^\prime = l^\prime$ is a moot point here, since, as mentioned before, a single point $q^\prime = l^\prime$, being of measure $0$, does not contribute to the $q^\prime$-integral in Eq.~\ref{eq:JOneQInt}. 
\\
\\  
Our goal is to obtain, analytically, an expression for $J_1$ in terms of small DEM $l^\prime (l^\prime \ll 1)$. We will bypass evaluating the integrals in Eq.~\ref{eq:JOneQInt} in closed form, by Taylor-expanding them for small $l^\prime$. To that end Eq.~\ref{eq:JOneQInt} is first written in the following way.
\begin{equation}
\begin{aligned}
& J_1 = \frac{\pi}{2 {l^\prime}^2}\Bigg[ 2{l^\prime}^2\int_0^\infty dq^\prime e^{-2q^\prime \tilde{d}} \frac{1}{{q^\prime}^2+2} + \psi(l^\prime) - 2{l^\prime}^2\chi(l^\prime) \Bigg],
\label{eq:JOnePsiChi}
\end{aligned}
\end{equation}
where, 
\begin{subequations}
\begin{equation}
\psi(l^\prime) = \int_0^{l^\prime} dq^\prime e^{-2q^\prime \tilde{d}} \frac{2{q^\prime}^2}{{q^\prime}^2+2} \\
\label{eq:Psi}
\end{equation}
\begin{equation}
\chi(l^\prime) = \int_0^{l^\prime} dq^\prime \frac{e^{-2q^\prime \tilde{d}}}{{q^\prime}^2+2}
\label{eq:Chi}
\end{equation}
\label{eq:PsiChi}
\end{subequations}
\\
\\
Next $\psi(l^\prime)$ and $\chi(l^\prime)$ are Taylor-expanded for small $l^\prime$. First, $\psi(l^\prime)$ is expanded around the point $l^\prime=0$ as follows. $\psi(l^\prime) \approx \psi(0) + \frac{d\psi}{dl^\prime}|_{l^\prime=0}l^\prime + \frac{1}{2!}\frac{d^2\psi}{d{l^\prime}^2}|_{l^\prime=0}{l^\prime}^2 + \frac{1}{3!}\frac{d^3\psi}{d{l^\prime}^3}|_{l^\prime=0} {l^\prime}^3 + \frac{1}{4!}\frac{d^4\psi}{d{l^\prime}^4}|_{l^\prime=0}{l^\prime}^4+... $.  We truncate the Taylor expansion up to the $o({l^\prime}^4)$ term, so that, as per Eq.~\ref{eq:JOnePsiChi}, the factor $\frac{1}{{l^\prime}^2}$ multiplying $\psi(l^\prime)$, will produce terms upto $o({l^\prime}^2)$. It is our goal to keep the energy expansion up to $o({l^\prime}^2)$. This is in line with the polaron energy calculations in conventional polar crystals, in which the $o({l^\prime}^2)$ term gives the mass correction of the electron due to polaron formation. $\frac{d\psi}{dl^\prime}$ can be calculated from Eq.~\ref{eq:Psi} using the second fundamental theorem of integral calculus, as per which, $\frac{d\psi}{dl^\prime}$ is simply the integrand of Eq.~\ref{eq:Psi} with $q^\prime$ being replaced by the variable $l^\prime$. So,

\begin{equation}
\begin{aligned}
&\frac{d\psi}{dl^\prime}  =  e^{-2l^\prime \tilde{d}} \frac{2{l^\prime}^2}{{l^\prime}^2+2} 
\label{eq:PsiFristDeriv}
\end{aligned}
\end{equation}
\\
\\
Calculating $\frac{d^2\psi}{d{l^\prime}^2}, \frac{d^3\psi}{d{l^\prime}^3}, \frac{d^4\psi}{d{l^\prime}^4}$, etc. from Eq.~\ref{eq:PsiFristDeriv} is straightforward. The derivatives of $\psi$ are calculated up to the order 4, and are evaluated at $l^\prime=0$, to give, $\frac{d\psi}{dl^\prime}|_{l^\prime=0} = 0, \frac{d^2\psi}{d{l^\prime}^2}|_{l=0} = 0, \frac{d^3\psi}{d{l^\prime}^3}|_{l^\prime=0} = 2,$ and $\frac{d^4\psi}{d{l^\prime}^4}|_{l^\prime=0} = -12\tilde{d}$. Using these in the above-mentioned Taylor-expansion of $\psi(l^\prime)$ about $l^\prime=0$, one obtains
\begin{equation}
\begin{aligned}
&\psi(l^\prime)  \approx  \frac{1}{3}{l^\prime}^3 - \frac{\tilde{d}}{2}{l^\prime}^4 
\label{eq:PsiTaylor}
\end{aligned}
\end{equation}
\\
\\
In an exactly analogous way, $\chi(l^\prime)$ can be Taylor-expanded using the integral given by  Eq.~\ref{eq:Chi}. We simply mention the result in the following.
\begin{equation}
\begin{aligned}
&\chi(l^\prime)  \approx  \frac{1}{2}l^\prime - \frac{\tilde{d}}{2}{l^\prime}^2 
\label{eq:ChiTaylor}
\end{aligned}
\end{equation}
$\chi(l^\prime)$ has been expanded upto $o({l^\prime}^2)$, since there is already an ${l^\prime}^2$ term multiplied to it in Eq.~\ref{eq:JOnePsiChi}, and our goal is to keep the $J_1$ in Eq.~\ref{eq:JOnePsiChi} upto $o({l^\prime}^2)$. Using Eq.~\ref{eq:PsiTaylor} and Eq.~\ref{eq:ChiTaylor} in Eq.~\ref{eq:JOnePsiChi} one obtains

\begin{equation}
\begin{aligned}
J_1 \approx \pi \int_0^\infty dq^\prime \frac{e^{-2q^\prime\tilde{d}}}{2+{q^\prime}^2} 
-\frac{\pi}{3}l^\prime +\frac{\pi \tilde{d}}{4}{l^\prime}^2
\label{eq:JOneFinal}
\end{aligned}
\end{equation}
The above expression of $J_1$ shows up in Eq.~\ref{eq:JOneTwoFinal}.
\\
\\

\subsection{Evaluation of $J_2$}
Expressing $\sin \phi$ and $\cos \phi$ in terms of $z$ in $J_2$, as given by Eq.~\ref{eq:JTwoQPhi}, $J_2$ can be written in terms of the contour integral of the complex variable $z \equiv e^{i\phi}$ as follows
\begin{equation}
\begin{aligned}
& J_2 = \int_0^\infty dq^\prime \frac{{q^\prime}^2e^{-2q^\prime \tilde{d}}}{4(q^2+2)^2}\oint dz h(z),
\label{eq:JTwoQNContour}
\end{aligned}
\end{equation}

where 
\begin{equation}
\begin{aligned}
& h(z) = -i\frac{(z^2-1)^2(z^2+1)}{z^3(z^2-\frac{{l^\prime}^2+{q^\prime}^2}{l^\prime q^\prime}z+1)},
\label{eq:JTwoContour}
\end{aligned}
\end{equation}
\\
\\
The function $h(z)$ in Eq.~\ref{eq:JTwoContour} has a pole of order 3 at $z=0$. There are two other poles of h(z), of order 1, viz., $z_1$ and $z_2$, which are the same ones as given by Eq.~\ref{eq:ZOneZTwo}. Just like before, of all the three poles, only two, viz., $0$ and $z_2$ are inside the unit circle, as shown in Fig.~\ref{fig:ComplexUnitCir}. Hence the residues of $h(z)$ have to be evaluated only at $z = 0$, and $z = z_2$ in order to obtain $\oint dz h(z)$. $z_2$ being a simple pole, the residue of $h(z)$ will be given by $Res (h(z))|_{z=z_2} = \lim_{z \to z_2} (z-z_2)h(z)$.                
$z=0$, being a pole of order $3$, $Res (f(z))|_{z=0} = \lim_{z \to 0}\frac{1}{2}\frac{d^2}{dz^2}\Bigg[z^3f(z)\Bigg]$. Evaluating and then adding these residues together, after some algebra involving the repeated use of the identity $z_1z_2 = 1$, one obtains a rather simple result: $\sum_i (Res(f(z)))\Bigg|_{z = z_i} = -2i z_2^2$. This, along with Eq.~\ref{eq:ResThm}, gives the following expression for the contour integral.   

\begin{equation}
\begin{aligned}
\oint dz h(z) = 4\pi z_2^2 
\label{eq:HContourInt}
\end{aligned}
\end{equation}
\\
\\
Using Eq.~\ref{eq:HContourInt} in Eq.~\ref{eq:JTwoQNContour}, after having replaced $z_2$ as given in Eq.~\ref{eq:ZOneZTwo}, one obtains the following expression for $J_2$  
\begin{equation}
\begin{aligned}
& J_2 = \pi \Bigg[ \frac{1}{{l^\prime}^2}\int_0^{l^\prime} dq^\prime \frac{{q^\prime}^4e^{-2q^\prime \tilde{d}}}{({q^\prime}^2+2)^2} + {l^\prime}^2\int_{l^\prime}^\infty dq^\prime \frac{e^{-2q^\prime \tilde{d}}}{({q^\prime}^2+2)^2} \Bigg]
\label{eq:J2QInt}
\end{aligned}
\end{equation}
\\
\\
Eq.~\ref{eq:J2QInt} needs to be expanded for small $l^\prime$, as was done in case of $J_1$. To that end, Eq.~\ref{eq:J2QInt} is written in the following form first.
\begin{equation}
\begin{aligned}
& J_2 = \pi \Bigg[ {l^\prime}^2 \int_0^\infty dq^\prime\frac{e^{-2q^\prime \tilde{d}}}{({q^\prime}^2+2)^2} 
+ \frac{1}{{l^\prime}^2}\alpha(l^\prime) - {l^\prime}^2\beta(l^\prime)  \Bigg],
\label{eq:J2QIntAlphaBeta}
\end{aligned}
\end{equation}

where the functions $\alpha(l^\prime)$, and $\beta(l^\prime)$ are given by 

\begin{subequations}
\begin{equation}
\alpha(l^\prime) = \int_0^{l^\prime} dq^\prime \frac{{q^\prime}^4e^{-2q^\prime \tilde{d}}}{({q^\prime}^2+2)^2} \\
\label{eq:Alpha}
\end{equation}
\begin{equation}
\beta(l^\prime) = \int_0^{l^\prime} dq^\prime \frac{e^{-2q^\prime \tilde{d}}}{({q^\prime}^2+2)^2}
\label{eq:Beta}
\end{equation}
\label{eq:AlphaBeta}
\end{subequations}
\\
\\
We will next apply the Taylor expansion method to the integrals given by Eq.~\ref{eq:AlphaBeta}. Since, we are interested in terms up to $o({l^\prime}^2)$ in Eq.~\ref{eq:J2QIntAlphaBeta}, we need to expand $\alpha(l^\prime)$ upto $o({l^\prime}^4)$ and $\beta(l^\prime)$ upto $o({l^\prime}^0)$. As for $\alpha(l^\prime)$ given by Eq.~\ref{eq:Alpha}, using the second fundamental theorem of integral calculus, one can write $\frac{d\alpha(l^\prime)}{dl^\prime} = \frac{{l^\prime}^4e^{-2l^\prime \tilde{d}}}{({l^\prime}^2+2)^2}$. From this expression it is possible to compute derivatives of $\alpha(l^\prime)$ of orders up to infinity. Hence it is a matter of straightforward calculation to show that $\frac{d\alpha(l^\prime)}{dl^\prime}|_{l^\prime=0} = \frac{d^2\alpha(l^\prime)}{d{l^\prime}^2}|_{l^\prime=0} = \frac{d^3\alpha(l^\prime)}{d{l^\prime}^3}|_{l^\prime=0} = \frac{d^4\alpha(l^\prime)}{d{l^\prime}^4}|_{l^\prime=0} = 0$. Also, $\alpha(l^\prime)|_{l^\prime=0} = 0$, as is obvious from Eq.~\ref{eq:Alpha}. Hence Taylor expanding $\alpha(l^\prime)$ up to $o({l^\prime}^4)$, $\alpha(l^\prime) \approx 0$. Hence there is no contribution of $\alpha(l^\prime)$ to $J_2$ up to $o({l^\prime}^2)$. 
\\
\\   
As for $\beta(l^\prime)$, $\beta(l^\prime = 0) = 0$, as can be seen from the definition of $\beta$ as given by Eq.~\ref{eq:Beta}. 
For small $l^\prime$, the first leading order term in $\beta(l^\prime)$ is $o({l^\prime})$. Hence ${l^\prime}^2 \beta(l^\prime)$ term in Eq.~\ref{eq:J2QIntAlphaBeta} does not have any contribution up to $o({l^\prime}^2)$. Hence neither $\alpha(l^\prime)$, nor $\beta(l^\prime)$ contribute to $J_2$ in Eq.~\ref{eq:J2QIntAlphaBeta} up to the order of our interest. Hence following Eq.~\ref{eq:J2QIntAlphaBeta}, $o({l^\prime}^2)$ expansion of $J_2$  assumes the following form. 
\begin{equation}
\begin{aligned}
& J_2 \approx {l^\prime}^2 \pi \int_0^\infty dq^\prime\frac{e^{-2q^\prime \tilde{d}}}{({q^\prime}^2+2)^2}
\label{eq:J2Final}
\end{aligned}
\end{equation} 
This expression for $J_2$ shows up in Eq.~\ref{eq:JOneTwoFinal}. 

\subsection{Evaluation of $J_3$}
Finally, we will obtain an $o({l^\prime}^2)$ expansion for $J_3$, as given in Eq.~\ref{eq:JThreeQPhi}. To that end we will first carry out the $\phi^\prime$ integration by our standard technique of replacing the $\phi^\prime$ integration by the contour integration w.r.t  $z \equiv e^{i\phi^\prime}$. Replacing $\sin \phi^\prime$ by $\frac{1}{2i}(z-z^{-1})$ and $\cos \phi^\prime$ by $\frac{1}{2}(z+z^{-1})$ in the expression for $J_3$, and changing $d\phi^\prime$ to $\frac{dz}{iz}$, $J_3$ assumes the following form   

\begin{equation}
\begin{aligned}
& J_3 = l^\prime \int_0^\infty dq^\prime {q^\prime}^3\frac{e^{-2q^\prime \tilde{d}}}{4({q^\prime}^2+2)^3}\oint dz t(z),
\label{eq:J3QNContour}
\end{aligned}
\end{equation}

where 
\begin{equation}
\begin{aligned}
& t(z) = -i\frac{(z^4-1)^2}{z^4(z^2-\frac{{l^\prime}^2+{q^\prime}^2}{l^\prime q^\prime}z+1)},
\label{eq:J3Contour}
\end{aligned}
\end{equation}
\\
\\
The function $t(z)$ has a pole of order 4 at $z=0$. There are two other poles of order 1: $z_1$ and $z_2$ of $t(z)$, which are the same ones as given by Eq.~\ref{eq:ZOneZTwo}. As was the case for $J_1$ and $J_2$, of all the three poles only two, viz., $0$ and $z_2$ are inside the unit circle, as shown in Fig.~\ref{fig:ComplexUnitCir}. $z_2$ being a simple pole, the residue of $t(z)$ will be given by $Res (t(z))|_{z=z_2} = \lim_{z \to z_2} (z-z_2)t(z)$. $z=0$, being a pole of order $4$, $Res (t(z))|_{z=0} = \lim_{z \to 0}\frac{1}{3!}\frac{d^2}{dz^2}\Bigg[z^3t(z)\Bigg]$. Evaluating all of these residues and adding them together, after some algebra involving the use of the identity $z_1z_2=1$, one obtains: $\sum_i (Res(t(z)))\Bigg|_{z = z_i} = -2i z_2(1+{z_2}^2)$. Using this in the residue theorem of Eq.~\ref{eq:ResThm}, the following is derived.    

\begin{equation}
\begin{aligned}
\oint dz t(z) = 4\pi z_2(1+{z_2}^2) 
\label{eq:TContourInt}
\end{aligned}
\end{equation}
\\
\\
Using Eq.~\ref{eq:TContourInt} in Eq.~\ref{eq:J3QNContour}, after having replaced $z_2$ as given in Eq.~\ref{eq:ZOneZTwo}, one obtains the following expression for $J_3$  
\begin{equation}
\begin{aligned}
J_3 =& \pi \Bigg[ \int_0^{l^\prime} dq^\prime \Bigg(1+\frac{{q^\prime}^2}{{l^\prime}^2}\Bigg) \frac{{q^\prime}^4 e^{-2q^\prime \tilde{d}}}{({q^\prime}^2+2)^3}\\
& \hspace{15mm}+ \int_{l^\prime}^{\infty} dq^\prime \Bigg(1+\frac{{l^\prime}^2}{{q^\prime}^2}\Bigg) \frac{{q^\prime}^2{l^\prime}^2 e^{-2q^\prime \tilde{d}}}{({q^\prime}^2+2)^3}\Bigg]
\label{eq:J3QInt}
\end{aligned}
\end{equation}
$J_3$ can be expanded for small $l^\prime$, in exactly the same way as was done for $J_1$ and $J_2$. It can be shown that the first integral in Eq.~\ref{eq:J3QInt} does not have any contribution upto $o({l^\prime}^2)$. The first non-zero term of the second integral in Eq.~\ref{eq:J3QInt} can be shown to be of $o({l^\prime}^2)$, which will be sole contributor to $J_3$ upto  $o({l^\prime}^2)$. Evaluating that term, one obtains the following $o({l^\prime}^2)$ expression for $J_3$.

\begin{equation}
\begin{aligned}
& J_3 \approx {l^\prime}^2 \pi \int_0^\infty dq^\prime\frac{{q^\prime}^2 e^{-2q^\prime \tilde{d}}}{({q^\prime}^2+2)^3}
\label{eq:J3Final}
\end{aligned}
\end{equation} 
This expression for $J_3$ shows up in Eq.~\ref{eq:JOneTwoFinal}.

\section{Evaluation of the Deacy Rate for BL graphene electrons}\label{app:app2}
Having written $d^2q^\prime$ as $q^\prime dq^\prime d\phi^\prime$, $q^\prime$ integral in Eq.~\ref{eq:ScatterRateBLDimless} will be carried out first\cite{Feynman}. The argument of the $\delta$ function in Eq.~\ref{eq:ScatterRateBLDimless} is written as 
\begin{equation}
\begin{aligned}
f_1(q^\prime) \equiv \frac{{\vec{l^\prime}-\vec{q^\prime}}^2}{2} +1 - \frac{{\vec{l^\prime}}^2}{2}
\label{eq:DeltaArgFOne}
\end{aligned}
\end{equation}
Next the following identity involving a delta-function is used in Eq.~\ref{eq:ScatterRateBLDimless}. $\delta(f_1(q^\prime)) = \sum_{i} \frac{\delta (q^\prime-q_i)}{|\frac{df_1}{dq^\prime}|_{q^\prime = q_i}}$, where $q_i$'s are the zeros of the function $f_1(q^\prime)$.  Thus Eq.~\ref{eq:ScatterRateBLDimless} assumes the following form.

\begin{equation} 
\begin{aligned}
\frac{1}{\tau} = &\frac{\tilde{\epsilon}e^2(m\omega_s)^{\frac{1}{2}}}{\hbar^{\frac{3}{2}}} \int d\phi^\prime \int_0^\infty dq^\prime \sum_i e^{-2q^\prime_i \tilde{d}}
\\&\Bigg (1-\frac{q_i^2\sin^2 \phi^\prime}{{l^\prime}^2 - 2}\Bigg)\frac{\delta(q^\prime - q_i)}{\sqrt{{l^\prime}^2 \cos^2 \phi^\prime -2}}
\label{eq:BL DeltaQPhiInt}
\end{aligned}
\end{equation}
\\
\\
The $q_i$'s appearing in Eq.~\ref{eq:BL DeltaQPhiInt} are $q_{i =1,2} = l^\prime \cos\phi^\prime \pm \sqrt{{l^\prime}^2\cos^2 \phi^\prime-2}$, as obtained by solving the equation $f_1(q^\prime) = 0$, with $f_1(q^\prime)$ given by Eq.~\ref{eq:DeltaArgFOne}. From the expressions of $q_i$'s above, it is clear that $\cos \phi^\prime$ should be greater than $\frac{\sqrt{2}}{l^\prime}$ for the $q_i$'s to be real and positive. That sets an upper limit to the range of the variable $\phi^\prime$ in the integration w.r.t  $\phi^\prime$: instead of varying from $0$ to $2\pi$, $\phi^\prime$ will vary from $0$ to $\cos^{-1}\frac{\sqrt{2}}{l^\prime}$. This is accompanied with a multiplication of the integrand in Eq.~\ref{eq:BL DeltaQPhiInt} by a factor of $2$, owing to the fact that $\cos \phi^\prime$ crosses the constant value $\frac{\sqrt{2}}{l^\prime}$ twice, when it does, as $\phi^\prime$ varies from $0$ to $2\pi$. Using the above-mentioned $q_i$'s, as well as properties of $\delta$ function, the $q^\prime$ integral in Eq.~\ref{eq:BL DeltaQPhiInt} can be carried out to yield the following expression for the decay rate $\frac{1}{\tau}$ for BL graphene.  

\begin{equation} 
\begin{aligned}
&\frac{1}{\tau} = \\
&\frac{\tilde{\epsilon}e^2(m\omega_s)^{\frac{1}{2}}}{\hbar^{\frac{3}{2}}} \int_0^{\cos^{-1}\frac{\sqrt{2}}{l^\prime}} d\phi^\prime  
\frac{4e^{-2l \tilde{d} \cos \phi^\prime}}{L_1} \\
&\Bigg [\cosh (2L_1 \tilde{d}) 
-\frac{\sin^2 \phi^\prime}{{l^\prime}^2-2}
\Bigg( (2{l^\prime}^2 \cos^2 \phi^\prime -2)\cosh (2L_1\tilde{d}) \\
&\hspace{30mm}-2 L_1 l^\prime \sinh(2L_1\tilde{d}) \cos \phi^\prime \Bigg) \Bigg],
\label{eq:BL DeltaPhiInt}
\end{aligned}
\end{equation}

where
\begin{equation} 
\begin{aligned}
L_1 \equiv \sqrt{{l^\prime}^2 \cos^2 \phi^\prime-2}
\end{aligned}
\end{equation}
Eq.~\ref{eq:BL DeltaPhiInt} is identical to Eq.~\ref{eq:BL DeltaPhiIntNotAppend} except for the spin degeneracy factor $g_s$.

\section{Evaluation of the Decay Rate for SL graphene electrons}\label{app:app3}
In Eq.~\ref{eq:EnergyLossSLG} the argument of the $\delta$ function is written as 

\begin{equation}
\begin{aligned}
f_2(q^\prime) \equiv \sqrt{{l^\prime}^2+ {q^\prime}^2-2l^\prime q^\prime \cos \phi^{\prime}}+1-l^\prime 
\label{eq:DeltaArgFTwo}
\end{aligned}
\end{equation}
Using $\delta(f_2(q^\prime)) = \sum_{i} \frac{\delta (q^\prime-q_i)}{|\frac{df_2}{dq^\prime}|_{q^\prime = q_i}}$, where $q_i$'s are the zeros of $f_2(q^\prime)$, one can write Eq.~\ref{eq:EnergyLossSLG} as follows.

\begin{equation} 
\begin{aligned}
\frac{1}{\tau} = &\frac{\tilde{\epsilon}e^2\omega_s}{2\hbar v_F} \int d\phi^{\prime} \int_0^\infty dq^\prime \sum_i e^{-2q^\prime_i \tilde{d}} \\
&\Bigg ( 2l^\prime-1 - q_i \cos \phi^{\prime} \Bigg )\frac{\delta(q^\prime - q_i)}{\sqrt{{l^\prime}^2 \cos^2 \phi^{\prime} -(2l-1)}}
\label{eq:SL DeltaQPhiInt}
\end{aligned}
\end{equation}
\\
\\
The $q_i$'s appearing in Eq.~\ref{eq:SL DeltaQPhiInt} are $q_{i =1,2} = l^\prime \cos\phi^{\prime} \pm \sqrt{{l^\prime}^2\cos^2 \phi^{\prime}-(2l^\prime-1)}$, as obtained by solving the equation $f_2(q^\prime) = 0$, where $f_2(q^\prime)$ is given by Eq.~\ref{eq:DeltaArgFTwo}. From the above-mentioned expressions of $q_i$'s, it is clear that $\cos \phi^{\prime}$ should be greater than $\sqrt{\frac{2l^\prime-1}{{l^\prime}^2}}$, for the $q_i$'s to be real and positive. That sets an upper limit to the range of the variable $\phi^{\prime}$ in the integration w.r.t  $\phi^{\prime}$: instead of varying from $0$ to $2\pi$, $\phi^{\prime}$ will vary from $0$ to $\cos^{-1}\sqrt{\frac{2l^\prime-1}{{l^\prime}^2}}$. This is accompanied with a multiplication of the integrand in Eq.~\ref{eq:SL DeltaQPhiInt} by a factor of $2$, owing to the fact that $\cos \phi^\prime$ crosses the constant value $\sqrt{\frac{2l^\prime-1}{{l^\prime}^2}}$ twice, when it does, as $\phi^\prime$ varies from $0$ to $2\pi$. Using the above-mentioned expressions for $q_i$'s, as well as properties of $\delta$ function, the $q^\prime$ integral in Eq.~\ref{eq:SL DeltaQPhiInt} can be carried out to yield the following expression for the decay rate $\frac{1}{\tau}$ for SL graphene.

\begin{equation} 
\begin{aligned}
\frac{1}{\tau} = &\frac{\tilde{\epsilon}e^2\omega_s}{\hbar v_F} \int_0^{\cos^{-1}\sqrt{\frac{2l^\prime-1}{{l^\prime}^2}}} d\phi^{\prime} \\  
&\frac{2e^{-2l^{\prime} \tilde{d} \cos \phi^{\prime}}}{L_2} 
\Bigg [(2l^\prime -1) \cosh (2 L_2 \tilde{d}) \\
&-\cos \phi^{\prime} \Bigg(l^\prime \cos \phi^{\prime} \cosh (2 L_2 \tilde{d})
-L_2 \sinh (2 L_2 \tilde{d}) \Bigg) \Bigg],
\label{eq:SL DeltaPhiInt}
\end{aligned}
\end{equation}

where
\begin{equation} 
\begin{aligned}
L_2 \equiv \sqrt{{l^\prime}^2 \cos^2 \phi^{\prime}-(2l^\prime -1)}
\end{aligned}
\end{equation}
Eq.~\ref{eq:SL DeltaPhiInt} is identical to Eq.~\ref{eq:SL DeltaPhiIntNotAppend} except for the spin degeneracy factor $g_s$.

\section{Evaluation of the decay Rate for Semi-Dirac}\label{app:app4}
In Eq.~\ref{eq:ScatterEnergySD} we will first integrate out the $q^\prime_y$ variable. To that end we will treat the argument $R$ of the $\delta$ function, given by Eq.~\ref{eq:SDEnergyDenom}, as a function of the variable $q^\prime_y$. Making use of the identity $\delta(R(q^\prime_y)) = \sum_{i} \frac{\delta (q^\prime_y-q^i_y)}{|\frac{\partial R}{\partial q^\prime_y}|_{q^\prime_y = q^i_y}}$, where $q^i_y$'s are the zeros of the the above-mentioned function $R(q^\prime_y)$, one can write Eq.~\ref{eq:ScatterEnergySD} as follows. 

\begin{equation}
\begin{aligned}
\frac{1}{\tau} =& \frac{\tilde{\epsilon}e^2(m\omega_s)^\frac{1}{2}}{2\hbar^{\frac{3}{2}}}\int dq^\prime_x \\
&\frac{\sqrt{\frac{1}{4}{l^\prime_x}^4+{l^\prime_y}^2}-1}
{\Bigg[\Bigg (\sqrt{\frac{1}{4}{l^\prime_x}^4+{l^\prime_y}^2}-1\Bigg)^2-\frac{1}{4}(l^\prime_x - q^\prime_x)^4 \Bigg]^\frac{1}{2}}\\
& \sum_{i=1,2} \frac{e^{-2\tilde d_{SD}\sqrt{{q^i_y}^2+\kappa {q_x^\prime}^2}}}{\sqrt{{q^i_y}^2+\kappa {q_x^\prime}^2}}\\
&\Bigg(1+\cos(\atan \frac{2l_y^\prime}{{l_x^\prime}^2}-\atan \frac{2(l_y^\prime - q^i_y)}{(l_x^\prime - q_x^\prime)^2})\Bigg),
\label{eq:SdDeltaSumq_x}
\end{aligned}
\end{equation}
\\
\\
The $q^i_y$ appearing in Eq.~\ref{eq:SdDeltaSumq_x}, which can be found easily by solving for the equation $R(q^\prime_y) = 0$, are given by 
\begin{equation}
\begin{aligned}
q^{i=1,2}_y = l_y \pm \Bigg[\Bigg (\sqrt{\frac{1}{4}{l^\prime_x}^4+{l^\prime_y}^2}-1\Bigg)^2-\frac{1}{4}(l^\prime_x - q^\prime_x)^4 \Bigg]^\frac{1}{2}
\label{eq:SDRoots}
\end{aligned}
\end{equation}
In Eq.~\ref{eq:SDRoots} `$\pm$' corresponds to the two roots, $i=1, 2$. Eq.~\ref{eq:SDRoots} sets limits on the possible values of $q^\prime_x$, since the expression inside the third bracket on the right side of Eq.~\ref{eq:SDRoots} has to be greater than $0$, for $q^{i=1,2}_y$ to have any real solution. Mathematically,  $\Bigg[\Bigg (\sqrt{\frac{1}{4}{l^\prime_x}^4+{l^\prime_y}^2}-1\Bigg)^2-\frac{1}{4}(l^\prime_x - q^\prime_x)^4 \Bigg] > 0$, which restricts the values of $q^\prime_x$ as follows.

\begin{equation}
\begin{aligned}
& \abs{q^\prime_x -l^\prime_x}  <  \sqrt{2}\sqrt{\sqrt{\frac{1}{4}{l^\prime_x}^4+{l^\prime_y}^2}-1}
\label{eq:SDInEq}
\end{aligned}
\end{equation}
\\
\\
Using Eq.~\ref{eq:SDRoots} in Eq.~\ref{eq:SdDeltaSumq_x}, as well as using the appropriate limits of $q^\prime_x$ commensurate with the inequality~\ref{eq:SDInEq}, as the limits of integration in Eq.~\ref{eq:SdDeltaSumq_x}, one obtains the following expression for the decay rate for semi-Dirac. 
\begin{equation}
\begin{aligned}
&\frac{1}{\tau} =\frac{\tilde{\epsilon}e^2(m\omega_s)^\frac{1}{2}}{2\hbar^{\frac{3}{2}}}\\
&\bigintss_{l^\prime_x - \sqrt{2}\sqrt{ \sqrt{\frac{1}{4}{l^\prime_x}^4+{l^\prime_y}^2}-1}}^{l^\prime_x + \sqrt{2}\sqrt{\sqrt{\frac{1}{4}{l^\prime_x}^4+{l^\prime_y}^2}-1}} dq^\prime_x
\frac{\sqrt{\frac{1}{4}{l^\prime_x}^4+{l^\prime_y}^2}-1}{L_3} \\
& \Bigg [ \frac{e^{-2\tilde{d}\sqrt{Q+2l^\prime_y L_3}}}
{\sqrt{Q+2l^\prime_y L_3}}
\Bigg(1+\cos(\atan \frac{2l_y^\prime}{{l_x^\prime}^2}+\atan \frac{2 L_3}{(l_x^\prime - q_x^\prime)^2})\Bigg) \\
& + \frac{e^{-2\tilde{d}\sqrt{Q-2l^\prime_y L_3}}}
{\sqrt{Q-2l^\prime_y L_3}}
\Bigg(1+\cos(\atan \frac{2l_y^\prime}{{l_x^\prime}^2}-\atan \frac{2 L_3}{(l_x^\prime - q_x^\prime)^2})\Bigg)  \Bigg],
\label{eq:SDDeltaFreeScatter}
\end{aligned}
\end{equation}
 
where $L_3 \equiv \Bigg[ \Bigg(\sqrt{\frac{1}{4}{l^\prime_x}^4+{l^\prime_y}^2}-1\Bigg)^2-\frac{1}{4}(l^\prime_x-q^\prime_x)^4\Bigg]^\frac{1}{2}$, and $Q \equiv \kappa {q_x^\prime}^2 + {l^\prime_y}^2 + L^2_3$
\\
\\
Eq.~\ref{eq:SDDeltaFreeScatter} is identical to Eq.~\ref{eq:SDDeltaFreeScatterNotAppend} except for the spin degeneracy factor $g_s$.

\end{document}